\documentclass[a4paper,fleqn,usenatbib,useAMS]{mnras}

\usepackage{graphicx}	
\usepackage{multicol}        
\usepackage{longtable}
\usepackage{natbib}
\usepackage{wrapfig}
\usepackage{array}
\usepackage{hyperref}

\usepackage{bm}		
\usepackage{pdflscape}	

\usepackage{enumerate}
\usepackage[shortlabels]{enumitem}
\usepackage{soul}
\usepackage[dvipsnames]{xcolor}
\usepackage{relsize}
\usepackage{amsmath}	
\usepackage{amssymb}	

\usepackage[T1]{fontenc}
\usepackage{ae,aecompl}

\usepackage{newtxtext,newtxmath}

\title[Galaxy--jet--filament orientation]{On the relationship between the cosmic web and the alignment of galaxies and AGN jets}

\author[Jung et al.]{S. Lyla Jung$^{1}$\thanks{e-mail: \href{mailto:lyla.jung@physics.ox.ac.uk}{lyla.jung@physics.ox.ac.uk}}, I. H. Whittam$^{1}$, M. J. Jarvis$^{1, 2}$, C. L. Hale$^{1}$, M. N. Tudorache$^{1}$, T. Yasin$^{1}$
\\
$^{1}$ Astrophysics, Denys Wilkinson Building, Department of Physics, University of Oxford, Keble Road, Oxford OX1 3RH, United Kingdom\\
$^{2}$ Department of Physics and Astronomy, University of the Western Cape, Robert Sobukwe Road, 7535 Bellville, Cape Town, South Africa
}

\begin{document}
\maketitle

\begin{abstract}
The impact of active galactic nuclei (AGN) on the evolution of galaxies explains the steep decrease in the number density of the most massive galaxies in the Universe. However, the fueling of the AGN and the efficiency of this feedback largely depend on their environment.
We use data from the Low Frequency Array (LOFAR) Two-metre Sky Survey Data Release 2 (LoTSS DR2), the Dark Energy Spectroscopic Instrument (DESI) Legacy Imaging Surveys, and the Sloan Digital Sky Survey (SDSS) DR12 to make the first study of the orientations of radio jets and their optical counterpart in relation to the cosmic web environment. 
We find that close to filaments ($\lesssim 11 \,\rm Mpc$), galaxies tend to have their optical major axes aligned with the nearest filaments. On the other hand, radio jets, which are generally aligned perpendicularly to the optical major axis of the host galaxy, show more randomised orientations with respect to host galaxies within $\lesssim 8 \,\rm Mpc$ of filaments. These results support the scenario that massive galaxies in cosmic filaments grow by numerous mergers directed along the orientation of the filaments while experiencing chaotic accretion of gas onto the central black hole. The AGN-driven jets consequently have a strong impact preferentially along the minor axes of dark matter halos within filaments. We discuss the implications of these results for large-scale radio jet alignments, intrinsic alignments between galaxies, and the azimuthal anisotropy of the distribution of circumgalactic medium and anisotropic quenching.
\end{abstract}
\begin{keywords}
methods: observational -- galaxies: evolution -- galaxies: jets -- (cosmology:) large-scale structure of Universe 
\end{keywords}

\section{Introduction}

It has long been established that supermassive black holes (SMBH) reside at the core of possibly all massive galaxies (\citealt{Kormendy_1995}).
As the strong gravity field around a SMBH accretes surrounding interstellar material onto the black hole, large amounts of energy can be released producing active galactic nucleus (AGN) activity. 
It is thought that AGN accretion occurs in two different modes, resulting in two different classes of AGN. The first is `jet mode' AGN, which accretes inefficiently from pockets of cold gas in the hot gas halo and emits the bulk of its power as a radio jet. The second is the radiatively efficient `radiative mode' sources typical of optical or X-ray selected AGN, which accretes efficiently from cold gas and may or may not have a radio jet \citep[e.g.][]{HeckmanBest2014}.

Galaxy formation models invoke feedback from such AGN to curtail the amount of star formation in massive galaxies and match the observed galaxy stellar mass function \citep[e.g.][]{Bower2006,Croton2006,Bower2017, Adams2021, McLeod2021}. This feedback is often prescribed as two related processes that are implemented in recent cosmological-volume galaxy formation simulations, albeit with slightly different prescriptions \citep[e.g.][]{Schaye2015, Dubois2016,Dave2019}. First, the radiative feedback from the hard ionisation field emanating from the accretion disk provides a heating mechanism for the surrounding gas, preventing it from cooling and condensing into cold gas necessary for star formation; the second mechanism arises from the generation of jet outflows, which can drive gas away from the host galaxy via mechanical feedback \citep[e.g.][]{Heckman2024}. These AGN jets may also have an impact on galaxy groups and clusters \citep[e.g.][]{Fabian2012, McNamara2012} potentially stimulating or truncating star formation within galaxies in such environments (\citealt{RawlingsJarvis2004, Hatch2014}).

However, such a clear distinction between `radiative' and `mechanical' feedback modes is possibly not completely correct. Many AGN that have both bright optical nuclei and broad line regions, which are the hallmark of the radiative mode, also produce powerful radio emission, traditionally termed radio-loud quasars and radio galaxies \citep[see ][ for a review]{UrryPadovani1995}. 
Furthermore, although large samples of bright radio sources tend to exhibit a bimodal distribution in their Eddington-scaled accretion rate, 
\citep{BestHeckman2012, Mingo2014}, more recent work using deeper radio data suggests that this bimodal distribution disappears towards lower radio luminosities \citep{Whittam2018, Whittam2022}. The relationship between the accretion rate and the generation of jets responsible for the mechanical feedback in these systems is therefore a critical aspect in enhancing our understanding of the role of AGN in galaxy evolution.

Theoretical models of AGN jet formation suggest that the jets are launched perpendicular to the gas accretion flow onto the black hole. If the gas accretion has a preferred direction with respect to an AGN host galaxy, this may result in a preferred orientation for jets. 
The direction of radio jets with respect to their host galaxies has been explored observationally by comparing the position angles of extended radio sources and their optical counterparts. 
Although early studies reported mixed views on alignment (\citealt{Mackay_1971, Palimaka_1979, Valtonen_1983, Birkinshaw_1985, Sansom_1987}), recent investigations based on large statistical samples suggest that radio jets tend to align with the optical minor axis on both the kpc-scale (\citealt{Battye_2009, Zheng_2024}) and the pc-scale (\citealt{Gil_2024}). 
Of particular relevance to this paper, \citet{Zheng_2024} analyse radio sources from the Low Frequency Array (LOFAR) Two-metre Sky Survey Data Release 2 (LoTSS DR2; \citealt{Shimwell_2022}) and show that the tendency for the jet-galaxy alignment depends on various properties such as the radio luminosity, the physical size of jets, the stellar mass, and the shape of host galaxies. This finding indicates that how the jets are launched and propagated is closely linked with the evolution of their host galaxies.

At a larger intergalactic scale, a relevant yet inconclusive question is whether there is a specific angular scale or regions in the sky in which the orientations of adjacent radio jets are coherent. If such coherency is indeed present, it implies that the spin axes of black holes are aligned over several tens of Mpc and larger physical scales, which can be attributed to the large-scale structure of the Universe. 
Some studies support the large-scale alignment of jets at several to tens of degrees scales (\citealt{Taylor_2016, Contigiani_2017, Panwar_2020, Mandarakas_2021}). However, more recent investigations do not support a statistically significant correlation among adjacent radio galaxies in the 3D space identified using photometric or spectroscopic redshifts (\citealt{Osinga_2020, Simonte_2023}). The link between the large-scale intergalactic environment and radio jets, therefore, remains unclear.

The influence of the large-scale cosmic web environment on galaxy evolution appears to be more clearly reflected in the shape and kinematic properties of galaxies (e.g., \citealt{Zhang_2015, Hirv_2017,Kraljic2021, Tudorache2022, Barsanti_2022, Barsanti_2023, Lee_2023}). 
Numerical simulations of cosmological structure formation demonstrate that as the universe evolves, galaxies and dark matter halos in filaments migrate towards denser nodes where galaxy groups and clusters develop (e.g., \citealt{Springel_2005}). Some of these galaxies and halos undergo mergers that predominantly occur along the direction of the filament they reside in (\citealt{Libeskind_2014, Kang_2015}). As a result, these simulations predict that massive galaxies in cosmic filaments tend to elongate along the filaments and have angular momentum perpendicular to the filament direction (\citealt{Aragon-Calvo_2007, Codis_2012, Libeskind_2013, Dubois_2014, Wang_2017}).

Observationally, \citet{Tempel_2013} report that elliptical galaxies in the SDSS DR8 spectroscopic galaxy sample (\citealt{ Aihara_2011, Tempel_2012}) preferentially have the optical minor axes perpendicular to the nearest cosmic filaments. On the other hand, lower-mass galaxies tend to have their spin axes aligned with the filaments \citep[e.g.][]{Kraljic2021, Tudorache2022}.
Such a non-negligible correlation between galaxy shape and cosmic filaments provides a physical basis for the intrinsic alignment among galaxies over large angular scales that affects the interpretation of weak-gravitational lensing-based cosmic shear measurements (\citealt{Hirata_2007, Joachimi_2011, Singh_2015, Chisari2015}).

In this paper, we explore the alignments between optical galaxies, radio jets, and cosmic filaments. Specifically, we focus on how the galaxy-filament and galaxy-jet alignments vary according to the cosmic web environment. The paper is structured as follows. Section \ref{sec:data} gives an overview of the data used in this study and explains how we define our samples. We present our analysis of the alignment between optical galaxies and cosmic filaments in Section \ref{sec:filament-galaxy}. The analysis of the alignment between optical galaxies and radio jets is presented in Section \ref{sec:jet-galaxy}. In Sections \ref{sec:mechanism} and \ref{sec:implication}, we discuss the physical mechanisms and implications of our findings. 
Section \ref{sec:summary} is the summary of this paper. 
Throughout the paper, we assume a $\Lambda$CDM cosmology based on \citet{Planck_2016} results: $\Omega_{\rm m} = 0.309$, $\Omega_{\rm \Lambda} = 0.691$, $\Omega_{\rm b} = 0.0486$, $H_{\rm 0} = 67.8\,\rm km\,s^{-1}Mpc^{-1}$, and $\sigma_{\rm 8} = 0.82$.

\section{Data}\label{sec:data}

\begin{figure*}
    \centering
    \includegraphics[width=\textwidth]{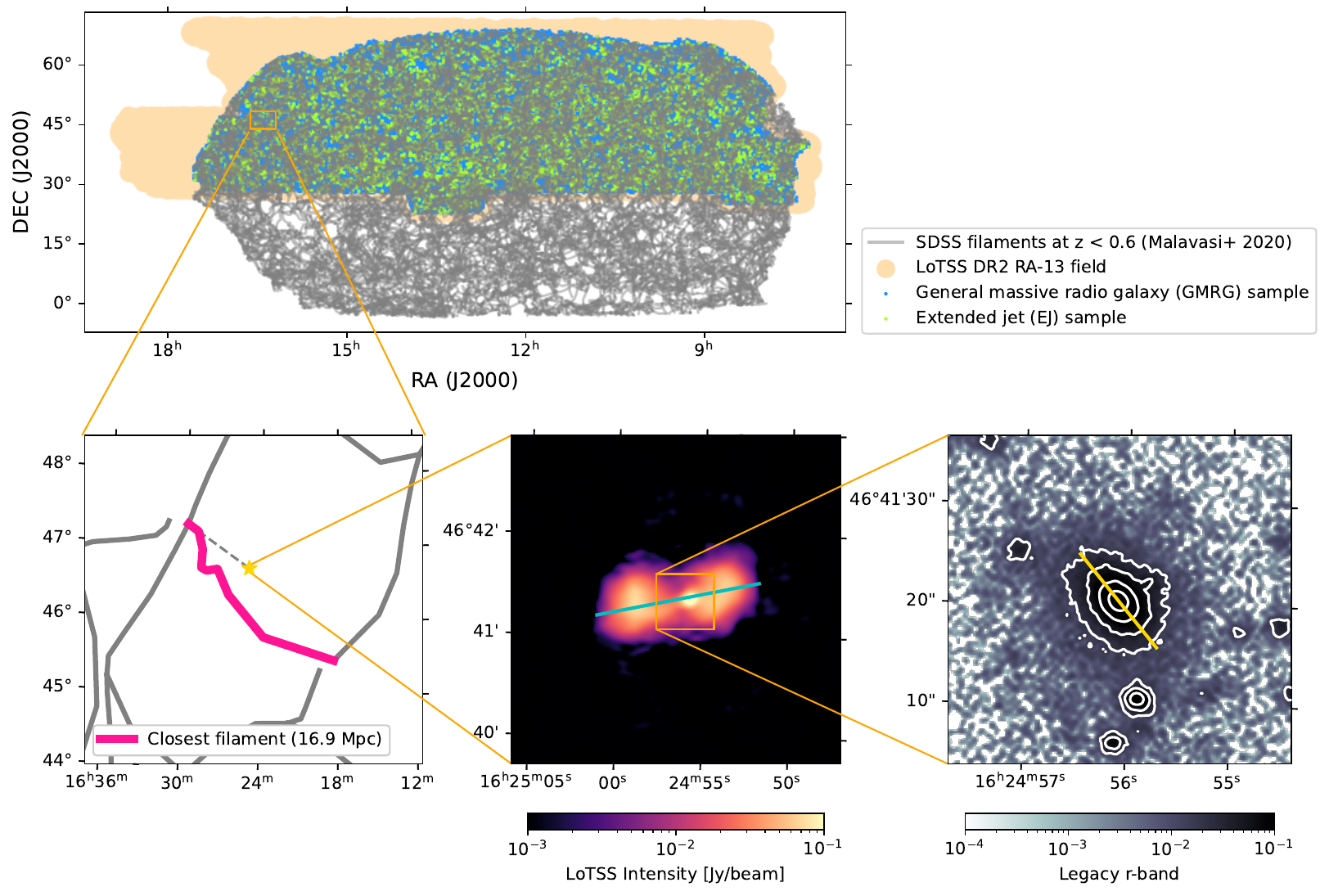}
    \caption{Top panel: The sky coverage of different data used in this study. The grey lines are cosmic filaments below redshift 0.6 catalogued by \citet{Malavasi_2020}. The peach colour shade shows the RA-13 field of LoTSS DR2. The general massive radio galaxy (GMRG) and the extended jet (EJ) samples defined in Section \ref{sec:sample} are shown with the blue and the green dots, respectively. 
    Bottom panels: we randomly select a galaxy in the EJ sample and show the surrounding cosmic filament distribution (left panel), the radio image (middle panel), and the $r$-band image (right panel). The filament closest to the example source, the radio major axis, and the optical major axis are shown in coloured lines (pink, blue, and yellow, respectively). The distance between the closest filament and the source is $16.9\,\rm Mpc$, as indicated in the legend in the left panel.
    }  
    \label{fig:illustration}
\end{figure*}

In this study, we compile a radio galaxy sample and their host galaxy properties and cosmic environments from several multiwavelength surveys. 
In Section \ref{sec:lotss}, we provide a brief overview of LoTSS DR2 (\citealt{Shimwell_2022}) and the radio-optical/infrared cross-match catalogue of LoTSS DR2 sources (\citealt{Hardcastle_2023}). 
Section \ref{sec:desi} describes the Dark Energy Spectroscopic Instrument (DESI) Legacy Imaging Surveys (\citealt{Dey_2019}) where we retrieve optical properties of our radio galaxy sample. 
For characterising the cosmic environment, we use the cosmic filament catalogue published by \citet{Malavasi_2020}, which will be explained in Section \ref{sec:filament}.
We describe the galaxy samples used in this study in Section \ref{sec:sample}.

\subsection{LoTSS DR2 and the optical value-added catalogue}\label{sec:lotss}

LoTSS DR2 covers 27\% of the northern sky ($\sim5700\,\rm deg^{2}$), split into two continuous fields, RA-1 and RA-13 centred at $\rm (RA, Dec.) = (1^{h}00^{m}00^{s}, +28^{\circ}00'00'')$ and $\rm (12^{h}45^{m}00^{s}, +44^{\circ}30'00'')$, respectively. 
LoTSS-DR2 observes the sky using LOFAR's (\citealt{vanHaarlem_2013}) High Band antennas (HBA), operating at $120-168\,\rm MHz$.
In total, about 4 million radio sources were identified throughout the fields using the Python Blob Detector and Source Finder ({\sc pybdsf}; \citealt{Mohan_2015}). 

The cross-matching between the radio and optical/infrared sources was carried out by \citet{Hardcastle_2023} using a combination of a likelihood-ratio method (\citealt{Sutherland_1992,McAlpine2012}) and visual inspection by citizen scientists via the Zooniverse project, `Radio Galaxy Zoo: LOFAR'\footnote{\url{http://lofargalaxyzoo.nl/}}, hereafter, RGZ(L). This strategy is similar to that described in \citet{Williams_2019} however, notably, \citet{Hardcastle_2023} restricted the sources which could go through the Galaxy Zoo process to those sources $\geq 4 \,\rm mJy$. This was due to the increased area of LoTSS DR2 compared to DR1 ($>10$ times of \citealt{Shimwell_2019}) leading to a significant increase in sources. Accurate positional information is crucial for WEAVE-LOFAR (\citealt{Smith_2016}), for which the wide-area survey has a minimum flux density limit of $8\,\rm mJy$. Therefore, $4\,\rm mJy$ was used so that multi-component sources whose total flux density would sum to $8\,\rm mJy$ could be identified and cross-matched together. 
This radio-optical cross-matched catalogue provides key parameters for this study, including the redshift and stellar mass of optical galaxies hosting the radio sources. For reference, of the 4 million radio sources in the catalogue, about 60 per cent (50 per cent) have redshift (stellar mass) measurements.

One of the key radio properties investigated in this study is the position angle of extended radio jets. 
We provide more details of how we identified the sample with extended radio jets in Section \ref{sec:sample}. Here, we explain how the position angle of radio sources is defined depending on the source structure (`S\_Code'). 

{\sc pybdsf} classifies the radio sources into three categories depending on the source structure. 
`S' sources are isolated and fitted by a single Gaussian. `C' sources are fitted by a single Gaussian and located within a group of emissions (i.e., an island) containing other sources. `M' sources are fitted with multiple Gaussians.
The `S\_Code' column in the cross-matched catalogue contains information about this {\sc pybdsf} source category or an additional category `Z' for sources identified as a composite source in the RGZ(L) where multiple {\sc pybdsf} sources needed associating
together into a single object.

For `S' and `M' type sources, we use the deconvolved position angle identified by {\sc pybdsf} (the `DC\_PA' column in the cross-matched catalogue). This is the angle of the source's major axis identified using image moment analysis\footnote{\url{https://pybdsf.readthedocs.io/en/latest/algorithms.html\#grouping-of-gaussians-into-sources}}. Note all position angles in this study are measured east of north. 
For `Z' type composite sources, the position angle is determined as the angle of the longest axis of a convex hull enclosing all components of a source (the `Composite\_PA' column in the cross-matched catalogue). 
For example, the lower middle panel in Fig. \ref{fig:illustration} shows a LoTSS image of a randomly selected radio source from the extended jet sample that we define shortly. The major axis of the source is shown in a blue straight line. The position angle of this source is $101^{\circ}$.

\subsection{DESI Legacy Imaging Surveys}\label{sec:desi}

While the LoTSS DR2 radio-optical cross-matching catalogue includes basic properties of radio galaxy hosts (e.g., coordinate, redshift, $g, r, z$-band magnitudes), we revisit the `sweep' catalogues of the DESI Legacy Imaging Surveys (\citealt{Dey_2019}) to obtain further information about the optical sources matched with the LoTSS radio sources. 
For this task, we use the `RELEASE', `BRICKID', and `OBJID' fields in the LoTSS catalogue, indicating the Legacy release number, brick ID, and object ID, respectively. 
All matched optical sources are from the Legacy Surveys DR8. At declination $\delta>+32^{\circ}.375$, the Legacy catalogue includes sources from the Beijing-Arizona Sky Survey (BASS; \citealt{Zou_2017}) and the Mayall z-band Legacy Survey (MzLS). At lower declinations, sources are taken from the DECam Legacy Survey (DECaLS).

The Legacy Survey performs photometry using the {\sc Tractor} algorithm (\citealt{Lang_2016}). 
In this framework, sources are classified into different types depending on the best-fitting morphological model (the `type' column in the DESI Legacy catalogue). As we are interested in measuring the position angle of optical galaxies, types relevant to our analysis are spatially extended sources: `EXP', `DEV', and `COMP'. These types indicate that the isophots of sources are best modelled with the exponential profile (`EXP'), deVaucouleurs profile (`DEV'), and the composite profile (exponential plus deVaucouleurs; `COMP'), respectively. 

Some of the key fields in the DESI Legacy catalogue we use for this study are the ellipticities of the optical sources (`SHAPEEXP\_E1', `SHAPEEXP\_E2', `SHAPEDEV\_E1', and `SHAPEDEV\_E2' columns). 
The optical ellipticity is defined as a complex number\footnote{This definition is taken from a weak gravitational lensing formalism (e.g., \citealt{Bridle_2009}). See \url{https://www.legacysurvey.org/dr8/catalogs/}.} using these fields:
\begin{equation}\label{eq:ellip_opt}
    \epsilon_{\rm opt} = \frac{a-b}{a+b}\exp(2i\,PA_{\rm opt})=\epsilon_{\rm 1}+i\epsilon_{\rm 2},
\end{equation}
where $a$, $b$, and $PA_{\rm opt}$ are the major and minor axes and the position angle of the ellipse fitted to the isophotes. Here, $\epsilon_{\rm 1}$ and $\epsilon_{\rm 2}$ are the two ellipticities given by the catalogue, each labelled with `\_E1' and `\_E2' in the columns. 
The above equation can be rearranged as
\begin{equation}
    PA_{\rm opt} = \frac{1}{2}\arctan\left(\frac{\epsilon_{\rm 2}}{\epsilon_{\rm 1}}\right).
\end{equation} 
For `EXP' and `DEV' type sources, the ellipticities $\epsilon_{\rm 1}$ and $\epsilon_{\rm 2}$, and therefore $PA_{\rm opt}$, are only defined for the best-fitting model, either exponential or deVaucouleurs. For `COMP' type sources, ellipticities are given separately for the exponential and deVaucouleurs components of the composite fit (labelled with `SHAPEEXP' and `SHAPEDEV' in the columns). We confirm that $\lesssim4\%$ of our sample falls into this case. For these galaxies, we calculate the position angle of each component separately and take the average value as the representative $PA_{\rm opt}$ of the source. 

We calculate the uncertainty of the position angles ($PA_{\rm opt, err}$) using the inverse variance of the ellipticities given in the catalogue (columns `SHAPEEXP\_E1\_IVAR', `SHAPEEXP\_E2\_IVAR', `SHAPEDEV\_E1\_IVAR', and `SHAPEDEV\_E2\_IVAR'). We apply the upper limit on the optical position angle uncertainty ($PA_{\rm opt, err}<1^{\circ}$) to limit our analysis to galaxies with well-defined $PA_{\rm opt}$. This criterion based on $PA_{\rm opt, err}$ eliminates only $\lesssim0.1\%$ of the sample. 
The lower right panel in Fig. \ref{fig:illustration} shows the Legacy $g$-band cutout image of the example source and its optical major axis (yellow line). The optical position angle of this source is $38.90^{\circ}$ with the uncertainty of $0.01^{\circ}$.

\subsection{SDSS cosmic filaments}\label{sec:filament}

\citet{Malavasi_2020} presents catalogues of cosmic filaments identified by applying the Discrete Persistent Structure Extractor (DisPerSE; \citealt{Sousbie_2011}) algorithm to the Sloan Digital Sky Survey (SDSS) galaxy distribution. 
In this study, we use a filament catalogue created using the SDSS DR12 (\citealt{Alam_2015}) LOWZ+CMASS sample (see \citealt{Reid_2016} for the definition of the sample). This sample spans a broader redshift range ($z \approx 0-0.8$) compared to the other available option, i.e., SDSS DR7 main galaxy sample (\citealt{Strauss_2002}) which covers $z \approx 0-0.3$. The large redshift coverage offers a large 3D volume of the Universe to study numerous cosmic filaments.

There are several input parameters incorporated into the DisPerSE filament-finding algorithm that could influence the overall properties (e.g., shape and length) of catalogued filaments: (i) the number of density field smoothing cycles, (ii) the persistence threshold and (iii) the number of skeleton smoothing cycles. In short, smoothing the density field prior to the filament finding helps reduce noise in the galaxy distribution, which can otherwise result in unphysical structures. Increasing the persistence threshold effectively reduces the detection of less significant short branches of cosmic filaments. Smoothing of the skeletons removes any sharp edges of the identified filaments. See Section 4 of \citet{Malavasi_2020} for a detailed discussion on the choice of parameters. We use the catalogue with one cycle of density smoothing, the persistence threshold of $3\sigma$, and one cycle of skeleton smoothing. The distribution of the filaments in the 2D sky plane is shown in the upper panel of Fig. \ref{fig:illustration} with grey lines.

Each DisPerSE filament is composed of multiple sampling points. The filament catalogue provides the 3D coordinates (RA, Dec., redshift) of sampling points. 
For the purpose of this study, we identify the cosmic filament closest to each galaxy in our samples and measure the position angle of the matched filament segment.  
We use {\tt astropy.coordinates.match\_coordinates\_3d} function to identify a filament sampling point closest to each LoTSS source with redshift measurements and measure the 3D distance between the matched filament and the source. In this process, all galaxy and filament coordinates are converted to Cartesian coordinates $(x, y, z)$ using 
\begin{equation}
\begin{split} 
&x = d\sin(\alpha)\cos(\beta)\\
&y = d\sin(\alpha)\sin(\beta)\\
&z = d\cos(\alpha),
\end{split}
\end{equation}
where $\alpha = \pi/2-\mathrm{Dec.}$, $\beta = \mathrm{RA}$, and $d$ is the distance to the filament sampling point calculated from its redshift.
The average spacing between adjacent DisPerSE filament sampling points is small ($\approx 14\,\rm Mpc$) compared to the average distance between a galaxy and the closest filament ($\approx 40\,\rm Mpc$; see Section \ref{sec:sample} for further discussion).
The bottom left panel of Fig. \ref{fig:illustration} illustrates the cosmic filament distribution close to the example galaxy (star symbol) shown in the right panels. The closest filament is highlighted with a thick pink-coloured line, and the dashed line connects the galaxy and the closest filament sampling point. Note that we use the distance in the 3D coordinate system to identify the closest filament sampling point, not the angular separation.

\begin{figure}
    \centering
    \includegraphics[width=\columnwidth]{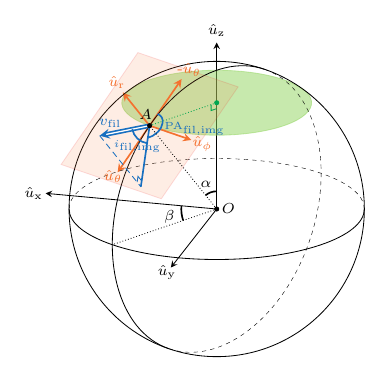}
    \caption{An illustration showing the Cartesian coordinate $(\hat{\textbf{\textit{u}}}_{\rm x}, \hat{\textbf{\textit{u}}}_{\rm y}, \hat{\textbf{\textit{u}}}_{\rm z})$ and the local spherical coordinate $(\hat{\textbf{\textit{u}}}_{\rm r}, \hat{\textbf{\textit{u}}}_{\rm \theta}, \hat{\textbf{\textit{u}}}_{\rm \phi})$ on the celestial sphere. The point $A$ and the vector $\textbf{\textit{v}}_{\rm fil}$ represent the location of a filament sampling point and the filament orientation vector, respectively. We measure the inclination and the position angle of the filament at point $A$ by transforming $\textbf{\textit{v}}_{\rm fil}$ from the Cartesian coordinate to the spherical coordinate. 
    See Section \ref{sec:filament} for further explanations.
    }  
    \label{fig:fil_pa_incl}
\end{figure}

Once the closest filament sampling point is identified, we define a filament orientation vector ($\textbf{\textit{v}}_{\rm fil}$) connecting the two adjacent sampling points of the same filament. If the closest sampling point is at the end of a filament, we use a vector connecting the closest point and the adjacent point of the same filament. 
As DisPerSE filaments are defined in the 3D space, this vector is also in the 3D Cartesian coordinate:
\begin{equation}
    \textbf{\textit{v}}_{\rm fil} = v_{\rm x} \hat{\textbf{\textit{u}}}_{\rm x}+ v_{\rm y} \hat{\textbf{\textit{u}}}_{\rm y}+ v_{\rm z} \hat{\textbf{\textit{u}}}_{\rm z},
\end{equation}
where $(v_{\rm x}, v_{\rm y}, v_{\rm z})$ are vector components in the Cartesian coordinate system with the basis of unit vectors $(\hat{\textbf{\textit{u}}}_{\rm x}, \hat{\textbf{\textit{u}}}_{\rm y}, \hat{\textbf{\textit{u}}}_{\rm z})$. 

Fig. \ref{fig:fil_pa_incl} illustrates how we convert a vector ($\textbf{\textit{v}}_{\rm fil}$) in the Cartesian coordinate to the local spherical coordinate with the basis of $(\hat{\textbf{\textit{u}}}_{\rm r}, \hat{\textbf{\textit{u}}}_{\rm \theta}, \hat{\textbf{\textit{u}}}_{\rm \phi})$ in order to measure the position angle and inclination of the vector. We set the location of a filament sampling point on the celestial sphere as $A$ and show an arbitrary filament orientation vector at this location with a blue double arrow. 
We convert the vector to the local spherical coordinate system using the following relation (\citealt{Lee_2007}):
\begin{equation}
    \begin{bmatrix}
    v_{\rm r} \\
    v_{\rm \theta} \\
    v_{\rm \phi}
    \end{bmatrix}
    =
    \begin{bmatrix}
    \sin{\alpha}\cos{\beta}& \cos{\alpha}\cos{\beta} & -\sin{\beta} \\
    \sin{\alpha} \sin{\beta} & \cos{\alpha} \sin{\beta} & \cos{\beta}\\
    \cos{\alpha} & -\sin{\alpha} & 0\\
    \end{bmatrix}^{-1}
    \begin{bmatrix}
    v_{\rm x} \\
    v_{\rm y} \\
    v_{\rm z}
    \end{bmatrix},
\end{equation}
where $(v_{\rm r}, v_{\rm \theta}, v_{\rm \phi})$ are components of the filament vector in the local spherical coordinate.
By definition, $(-\hat{\textbf{\textit{u}}}_{\rm \theta}, \hat{\textbf{\textit{u}}}_{\rm \phi})$ is the sky image plane local to the point $A$ (orange rectangle plane in Fig. \ref{fig:fil_pa_incl}) and $\hat{\textbf{\textit{u}}}_{\rm r}$ is the normal vector of this plane.
The inclination of the filament vector is the angle between $\textbf{\textit{v}}_{\rm fil}$ and the image plane. The position angle is the angle between $-\hat{\textbf{\textit{u}}}_{\rm \theta}$, i.e., the local north vector, and the $\textbf{\textit{v}}_{\rm fil}$ projected onto the image plane.
\begin{equation}
\begin{split}
    &i_{\rm fil, img} = \arcsin{\left(\frac{v_{\rm r}}{\sqrt{v_{\rm r}^{2}+v_{\rm \theta}^{2}+v_{\rm \phi}^{2}}}\right)}\\&\\
    &PA_{\rm fil, img} = \arctan{\left(-\frac{v_{\rm \phi}}{v_{\rm \theta}}\right)}
\end{split}
\end{equation}
Note that this position angle and inclination (as indicated by the notation ``$_{\rm img}$'') are specific to the image plane of the point $A$. This is because the coordinate basis $\hat{\textbf{\textit{u}}}_{\rm \phi}$ is defined as a tangent vector of the equal declination line of the celestial sphere (green circle parallel to the $x$-$y$ plane in Fig. \ref{fig:fil_pa_incl}) and, therefore, its vector magnitude $|\hat{\textbf{\textit{u}}}_{\rm \phi}|$ is dependent on the declination of the point $A$.  
In contrast, $|\hat{\textbf{\textit{u}}}_{\rm r}|$ and $|\hat{\textbf{\textit{u}}}_{\rm \theta}|$ do not vary with changing locations on the sphere. 
In light of this, we use the generalised inclination and position angle in our analysis.
\begin{equation}
\begin{split}
    &i_{\rm fil} = \arcsin{\left(\frac{v_{\rm r}}{\sqrt{v_{\rm r}^{2}+v_{\rm \theta}^{2}+\left(\frac{v_{\rm \phi}}{\cos(\mathrm{Dec.})}\right)^{2}}}\right)}\\
    &\\
    &PA_{\rm fil} = \arctan{\left(-\frac{v_{\rm \phi}}{v_{\rm \theta}\cos{(\mathrm{Dec.})}}\right)}
\end{split}
\end{equation}

\subsection{Sample selection}\label{sec:sample}

To study the alignment of optical galaxies, radio jets, and cosmic filaments, we select radio sources from the LoTSS optical cross-matched catalogue located within the cosmic volume covered by the SDSS filament catalogue. This means that, while the full LoTSS data come from two fields (RA-1 and RA-13), we limit our samples to sources in the RA-13 field (the peach-coloured shade in the upper panel of Fig. \ref{fig:illustration}), specifically sources less than $1^{\circ}$ angular distance away from any SDSS filament on the projected sky. This ensures any LoTSS sources outside the (RA, Dec.) coverage of the SDSS filaments are excluded from the analysis. 

Accurate redshift measurements of LoTSS sources are crucial in determining the location of radio sources with respect to cosmic filaments in 3D space. 
For sources in the LoTSS catalogue, both photometric and spectroscopic redshifts are provided, depending on availability. The `z\_best' and `z\_source' columns provide the best available redshift estimate and the origin of the best estimate, respectively. For the purpose of this study, we select sources with spectroscopic redshift measurement, either from the SDSS DR16 (\citealt{Ahumada_2020}), the DESI spectroscopic survey (\citealt{Desi_2024}), and the HETDEX data release (\citealt{Mentuch_2023}).
We limit the redshift of the LoTSS sources to $z<0.6$ where the number density of the SDSS DR12 LOWZ+CMASS sample within the survey volume remains flat (see Fig. 4 of \citealt{Malavasi_2020}). Beyond $z>0.6$, the filament identification is incomplete due to the limited sampling of the LOWZ+CMASS galaxy density field. 
Exclusively using sources with a spectroscopic redshift below 0.6 is crucial for our analysis as the accurate characterisation of the 3D locations of radio sources and cosmic filaments is key to quantifying the cosmic environment in which the sources reside. 
As mentioned earlier, 60 per cent of the sources in the LoTSS DR2 radio-optical cross-matching catalogue have redshift measurements and in the redshift range of $z<0.6$. About 30 per cent of these redshift estimates are spectroscopic and the rest are photometric. 
This sample selection does not affect the results that will follow, since the availability of a spectroscopic redshift does not depend on key parameters probed in this study, namely the distance between a radio source and the closest cosmic filament defined by SDSS, the orientation of radio jets, and the orientation of the optical host galaxy.

Two samples are configured for different scientific purposes.
The first is the general massive radio galaxy (GMRG) sample (blue markers in the top panel of Fig. \ref{fig:illustration}, though most overlap with the green markers that correspond to the second sample explained in the paragraph below). The GMRG sample consists of radio sources matched with an extended optical galaxy (i.e., a galaxy with a measurable optical position angle) with a stellar mass above $>10^{11}\,\rm M_{\rm \odot}$. In Section \ref{sec:filament-galaxy}, we use this sample to investigate the alignment of the galaxy's optical major axis and the cosmic filament orientation. 
There are 84409 sources that satisfy the selection criteria for the GMRG sample.

Secondly, we define the extended jet (EJ) sample as a subset of the GMRG sample that shows bright, extended (i.e., sources with a measurable radio position angle) radio jets (green markers in Fig. \ref{fig:illustration}). 
We use the EJ sample to investigate the alignment between radio jets and their host galaxies in Section \ref{sec:jet-galaxy}. 
We impose selection criteria for the EJ sample based on the radio properties as follows. In total, there are 7297 sources matching the selection criteria. 
\begin{enumerate}[wide = 0pt]
    \item High signal-to-noise: `Peak\_flux'/`Isl\_rms' > 10, where `Peak\_flux' is the $144\,\rm MHz$ peak flux density and `Isl\_rms' is the local root-mean-square noise.
    \item High radio luminosity: `L\_144' > $10^{24} \,\rm W\,Hz^{-1}$, where `L\_144' is the radio luminosity estimated from the total flux density of the source assuming a spectral index of 0.7\footnote{The spectral index, $\alpha$ is related to flux density $S_\nu$ by $S_{\nu} \propto \nu^{-\alpha}$}.
    \item Large angular size: `LAS' > 20 arcsec and `Resolved' == True, where `LAS' is the angular size estimated using a method specified in the `LAS\_from' column. The `Resolved' column is based on the resolution criterion of \citet{Shimwell_2022} and ensures a reliable measurement of `LAS'.
    \item Reliable radio position angle measurement: For `S\_Code' = `M' sources, `E\_PA' < $5^{\circ}$, where `E\_PA' is the error in $PA_{\rm radio}$; for `S\_Code' = `Z' sources, `Blend\_prob' < 0.2 and ‘Other\_prob’ < 0.2, where `Blend\_prob' and ‘Other\_prob’ are the probabilities for the source being blended or problematic.
\end{enumerate}

\begin{figure}
    \centering
    \includegraphics[width=\columnwidth]{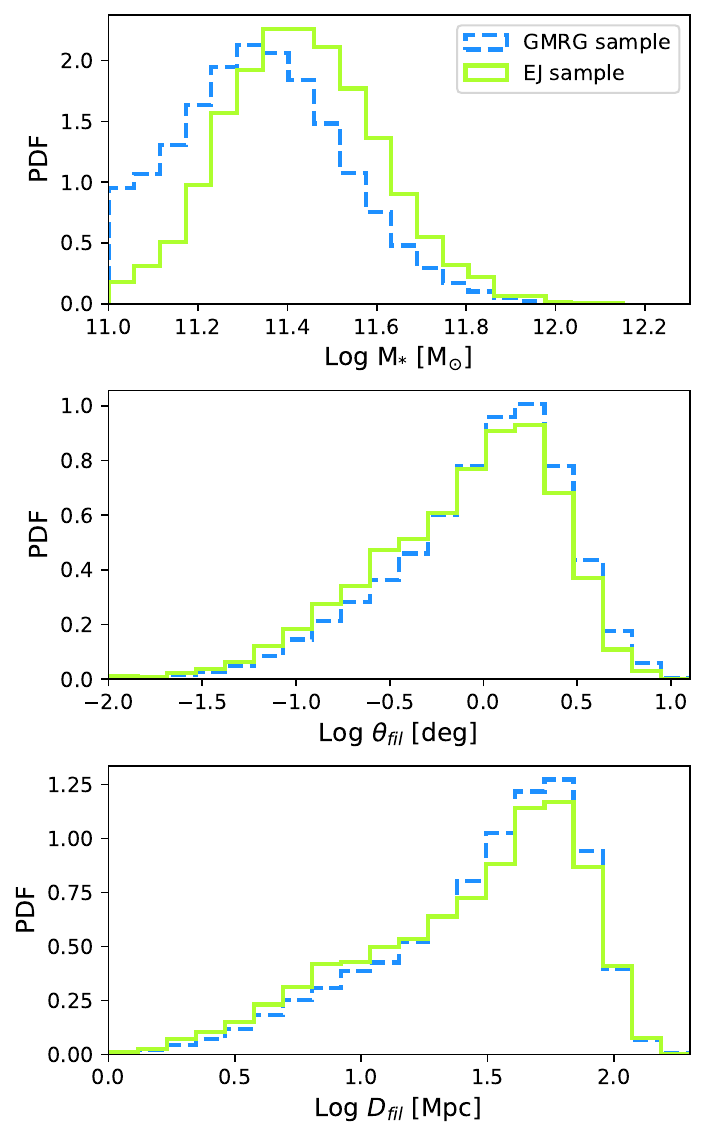}
    \caption{Histograms showing the stellar mass distribution (top panel) and the angular/physical distance from the closest filament distribution (middle/bottom panel) of the GMRG (blue dashed) and EJ (green) samples. Note that the EJ sample is a subset of the GMRG sample.
    }  
    \label{fig:dist_histogram}
\end{figure}

The top panel of Fig. \ref{fig:dist_histogram} shows the stellar mass distribution of the GMRG and EJ samples. All our sample galaxies are between $11<\log M_{*}/\rm M_{\odot}<12.33$. 
The stellar mass of the EJ sample is on average higher than the GMRG sample. 
This trend is well known, as powerful radio galaxies have long been associated with being hosted by the most massive galaxies, which contain the most massive black holes
(\citealt{Jarvis2001,McLure_2004, Herbert_2011, Whittam2022}). 

We also show histograms of the angular separation ($\theta_{\rm fil}$, middle panel) and the physical distance ($D_{\rm fil}$, bottom panel) between radio sources and their closest cosmic filament sampling point identified following the method described in Section \ref{sec:filament}. 
The long tails of the distributions towards the large $\theta_{\rm fil}$ and $D_{\rm fil}$ values include not only extremely isolated galaxies in the voids but also galaxies close to less prominent filaments that are not picked up by the DisPerSE filament finder. 
In the following sections, wherever we discuss the effects of filaments, we limit the samples to galaxies relatively close to the catalogued filaments (e.g., $D_{\rm fil}<80\,\rm Mpc$). 
By doing so, we omit galaxies that are close to unidentified filaments that potentially bear the imprint of filaments on their physical properties. Yet, there are already a sufficient number of galaxies close to the prominent filaments to obtain the statistical significance of the results presented in this paper.

\section{Results}\label{sec:result}

\subsection{Alignment between optical galaxy and cosmic filament}\label{sec:filament-galaxy}

In this section, we examine the alignment of optical galaxies with their nearest cosmic filaments using the GMRG sample. 

\subsubsection{Parallel transport method}\label{sec:parallel_transport}
\begin{figure}
    \centering
    \includegraphics[width=\columnwidth]{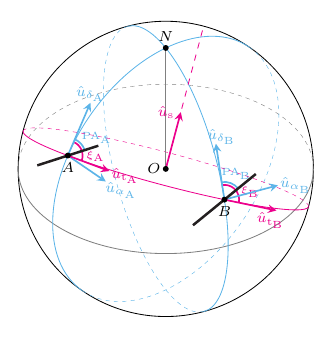}
    \caption{An illustration of parallel transport method. 
    In this illustration, there are two blue-coloured great circles connecting the North celestial pole ($N$) and the points $A$ and $B$, respectively.  
    Angles $PA_{\rm A}$ and $PA_{\rm B}$ are the position angles of the two black bars at each point, measured from the local north vectors ($\hat{\textbf{\textit{u}}}_{\rm \delta_{\rm A}}$ and $\hat{\textbf{\textit{u}}}_{\rm \delta_{\rm B}}$). The pink circle is the great circle connecting $A$ and $B$. Angles $\xi_{\rm A}$ and $\xi_{\rm B}$ are the angle between the local north vector and the tangent vector of the pink great circle ($\hat{\textbf{\textit{u}}}_{\rm t_{\rm A}}$ and $\hat{\textbf{\textit{u}}}_{\rm t_{\rm B}}$) at points $A$ and $B$, respectively. For further explanation of the parallel transport method, see the text in Section \ref{sec:parallel_transport}.
    }  
    \label{fig:parallel_transport}
\end{figure}

\begin{figure}
    \centering
    \includegraphics[width=\columnwidth]{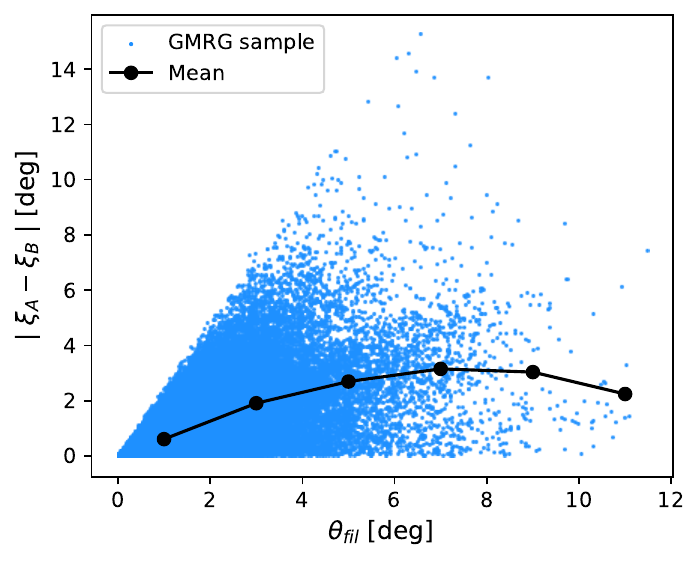}
    \caption{
    The degree of parallel transport applied to the GMRG sample as a function of the angular separation between the optical galaxy and the filament ($\theta_{\rm fil}$). Each blue marker corresponds to one GMRG sample galaxy. The black line shows the mean value of $\left|\xi_{\rm A}-\xi_{\rm B}\right|$ at a given $\theta_{\rm fil}$ window size of $2^{\circ}$.
    }  
    \label{fig:parallel_transport_degree}
\end{figure}

As shown in Fig. \ref{fig:dist_histogram}, a galaxy and the closest filament sampling point can be separated by significant angular distances in some cases. 
In this case, the position angles of the galaxy and the filament orientation should be compared with extra care. 
Fig. \ref{fig:parallel_transport} illustrates how we correctly compare two position angle measurements ($PA_{\rm A}$ and $PA_{\rm B}$) at two points on the celestial sphere, $A$ and $B$, using the parallel transport method (\citealt{Jain_2004, Contigiani_2017, Mandarakas_2021, Osinga_2020}). 
The blue great circles intersecting at point $N$ (north) are the local meridians at $A$ and $B$. 
The bold black bar at $A$ and $B$ represents the orientation of the object of interest. 
The position angle of a source is, by definition, measured from the axis pointing north along the local meridian.  
Therefore, the vector of the black bar at location $A$ can be expressed as
\begin{equation}
    \textbf{\textit{v}}_{\rm A} = \sin(PA_{\rm A}) \hat{\textbf{\textit{u}}}_{\rm \alpha_{\rm A}}+\cos(PA_{\rm A}) \hat{\textbf{\textit{u}}}_{\rm \delta_{\rm A}},
\end{equation}
where $\hat{\textbf{\textit{u}}}_{\rm \alpha_{\rm A}}$ and $\hat{\textbf{\textit{u}}}_{\rm \delta_{\rm A}}$ are the unit vectors local to the location $A$ pointing at north and east, respectively.
Similarly, the vector at point $B$ is written as 
\begin{equation}
    \textbf{\textit{v}}_{\rm B} = \sin(PA_{\rm B}) \hat{\textbf{\textit{u}}}_{\rm \alpha_{\rm B}}+\cos(PA_{\rm B}) \hat{\textbf{\textit{u}}}_{\rm \delta_{\rm B}}.
\end{equation}
Since the points $A$ and $B$ have non-negligible separation in the RA axis, the north vectors $\hat{\textbf{\textit{u}}}_{\rm \delta_{\rm A}}$ and $\hat{\textbf{\textit{u}}}_{\rm \delta_{\rm B}}$ point at different directions in the absolute reference frame. 
The parallel transport method exploits the fact that the angle between the orientation vectors, $\textbf{\textit{v}}_{\rm A}$ and $\textbf{\textit{v}}_{\rm B}$, and vectors tangential to the great circle connecting the two locations (the pink colour circle intersecting with $A$ and $B$) is invariant for parallel transport to any location along this great circle. 
The normal vector of this great circle is 
\begin{equation}
    \hat{\textbf{\textit{u}}}_{\rm s} = \frac{\hat{\textbf{\textit{u}}}_{\rm A}\times \hat{\textbf{\textit{u}}}_{\rm B}}{|\hat{\textbf{\textit{u}}}_{\rm A}\times \hat{\textbf{\textit{u}}}_{\rm B}|},
\end{equation}
where $\hat{\textbf{\textit{u}}}_{\rm A}$ and $\hat{\textbf{\textit{u}}}_{\rm B}$ are unit radial vectors pointing at points $A$ and $B$, respectively. 
The unit vectors tangential to the great circle at points $A$ and $B$ are
\begin{equation}
    \hat{\textbf{\textit{u}}}_{\rm t_{\rm A}} = \hat{\textbf{\textit{u}}}_{\rm s} \times \hat{\textbf{\textit{u}}}_{\rm A}
\end{equation}
and 
\begin{equation}
    \hat{\textbf{\textit{u}}}_{\rm t_{\rm B}} = \hat{\textbf{\textit{u}}}_{\rm s} \times \hat{\textbf{\textit{u}}}_{\rm B},
\end{equation}
respectively. 
The angle between the local north vectors and the tangent vectors can be expressed as 
\begin{equation}
    \xi_{\rm A} = \arccos{(\hat{\textbf{\textit{u}}}_{\rm \delta_{\rm A}}\cdot\hat{\textbf{\textit{u}}}_{\rm t_{\rm A}})}
\end{equation}
and
\begin{equation}
    \xi_{\rm B} = \arccos{(\hat{\textbf{\textit{u}}}_{\rm \delta_{\rm A}}\cdot\hat{\textbf{\textit{u}}}_{\rm t_{\rm B}})}.
\end{equation}
The angles $\xi_{\rm A}$ and $\xi_{\rm B}$ quantify the change of the local RA and Dec. basis with respect to the great circle. 
Transporting the orientation vector $\textbf{\textit{v}}_{\rm A}$ to the location $B$ makes a vector $\textbf{\textit{v}}'_{\rm A}$ with the following position angle with respect to the local meridian at $B$:
\begin{equation}
    PA'_{\rm A} = PA_{\rm A} - \xi_{\rm A} + \xi_{\rm B}.
\end{equation} 
Figure \ref{fig:parallel_transport_degree} shows the absolute degree of parallel transport, i.e., $\left|\xi_{\rm A}-\xi_{\rm B}\right|$, we apply when comparing the optical position angles of the GMRG sample (blue markers) to the position angle of the closest filament. The black line is the mean profile of the distribution. On average, parallel transport alters the position angles of the GMRG galaxies by $\lesssim3^{\circ}$. 

Finally, the dot product between the two orientation vectors leads to 
\begin{equation}
    \textbf{\textit{v}}'_{\rm A} \cdot \textbf{\textit{v}}_{\rm B} = \cos{(PA_{\rm A}-PA_{\rm B} - \xi_{\rm A} + \xi_{\rm B})},
\end{equation} 
i.e., the true angle between the two orientation vectors at points $A$ and $B$ is $PA_{\rm A}-PA_{\rm B} - \xi_{\rm A} + \xi_{\rm B}$. 
Note that, by definition, $PA_{\rm A}$ and $PA_{\rm B}$ range between $[0,180]^{\circ}$. 
When we refer to the angle between two orientations in the following paragraphs, we wrap the angles between $[0,90]^{\circ}$ as we do not define a direction for the filament or jet orientation.

\begin{figure*}
    \centering
    \includegraphics[width=0.95\textwidth]{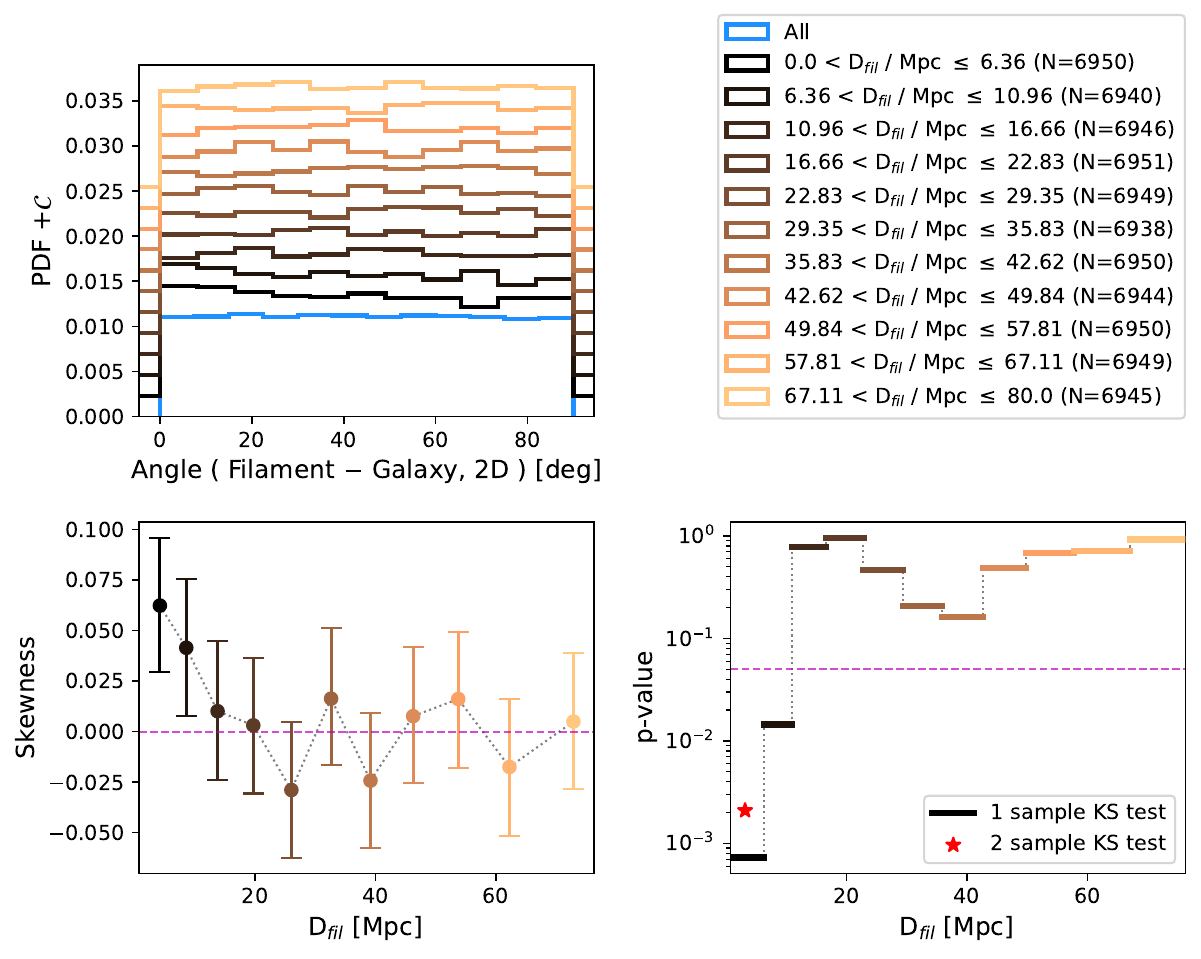}
    \caption{
    The distribution of the angle between the galaxy optical major axis and the closest filament orientation as a function of the distance to the filament. The GMRG sample is used for this analysis.
    Top panel: the histogram between $[0, 90]^{\circ}$. The blue line represents the distribution of all GMRG sample galaxies within $D_{\rm fil}<80\,\rm Mpc$. All other colour lines correspond to different $D_{\rm fil}$ bins. All histograms are vertically shifted, as denoted by $\mathcal{C}$ on the $y$-axis label, for better visibility. The $\mathcal{C}$ value is different for each histogram and can be determined by the $y$-axis value in the negative $x$-axis range.
    Bottom left panel: The skewness of the galaxy-filament angle distribution in different $D_{\rm fil}$ bins. The positive (negative) skewness indicates that the distribution is skewed towards $0^{\circ}$ ($90^{\circ}$). The horizontal dashed line shows where the skewness is zero.
    Bottom right panel: The thick horizontal lines show the p-values of one-sample KS tests comparing the uniform distribution and the filament-galaxy angle distribution in each bin. The red star symbol in the first bin shows the p-value of a two-sample KS test comparing the filament-galaxy angle distributions in $D_{\rm fil}\leq 6.36\,\rm Mpc$ and $D_{\rm fil}>6.36\,\rm Mpc$ ranges. The horizontal dashed line shows where the p-value is 0.05.
    }  
    \label{fig:filament-galaxy}
\end{figure*}

\begin{figure}
    \centering
    \includegraphics[width=\columnwidth]{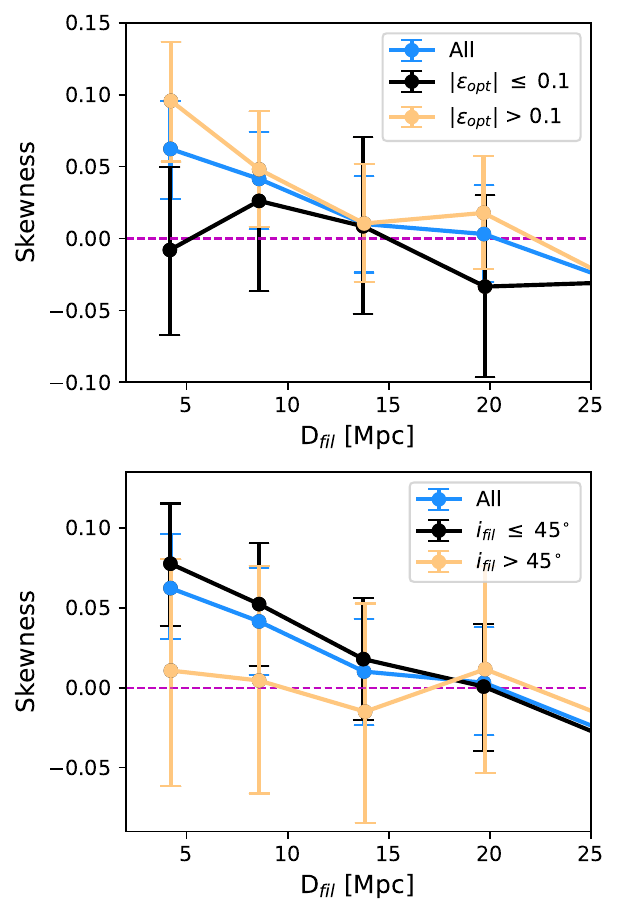}
    \caption{
    The skewness of the galaxy-filament angle distribution in different $D_{\rm fil}$ bins. In both panels, the blue line shows the skewness calculated using all galaxies in the GMRG sample, i.e., the result shown in the bottom left panel of Fig. \ref{fig:filament-galaxy}. 
    Top panel: The effect of varying optical ellipticity (black: $|\epsilon_{\rm opt}|\leq0.1$, yellow: $|\epsilon_{\rm opt}|>0.1$). Bottom panel: The effect of varying filament inclination (black: $i_{\rm fil}\leq45^{\circ}$, yellow: $i_{\rm fil}>45^{\circ}$).
    }  
    \label{fig:filament-galaxy_ellip_incl}
\end{figure}

\subsubsection{Results: galaxy-filament alignment in cosmic filaments}

The upper panel of Fig. \ref{fig:filament-galaxy} shows histograms of the angle between the galaxy major axis and the orientation of the closest filament. Angles close to zero indicate that galaxies are elongated along the direction of the filament. 
Note that a uniform distribution is expected for the angle between two randomly oriented 2D vectors on the sky plane. 
The blue line represents all GMRG sample galaxies within $D_{\rm fil} < 80\,\rm Mpc$. We find this distribution is consistent with being uniform between $0^{\circ}$ and $90^{\circ}$. 
The other lines are the histograms of the filament-galaxy angle in different ranges of $D_{\rm fil}$, shifted along the $y$-axis for better visibility. The sample is binned to contain roughly the same number of galaxies in each $D_{\rm fil}$ range. 
Although the distribution is nearly uniform in most binned samples, there is a subtle excess toward $0^{\circ}$ among galaxies close to the filaments ($D_{\rm fil}\leq6.36\,\rm Mpc$).

In order to quantify the excess, we calculate the skewness of each distribution defined as 
\begin{equation}
    G_{\rm 1} = \frac{\sqrt{N(N-1)}}{N-2}\frac{m_{\rm 3}}{m_{\rm 2}^{3/2}},
\end{equation}
i.e., the adjusted Fisher-Pearson coefficient, where 
\begin{equation}
    m_{\rm i} = \frac{1}{N}\mathlarger{\Sigma}^{N}_{i=1}(x_{\rm i}-\bar{x})^{i},
\end{equation}
$N$ is the total number of galaxies in each sample, and $\bar{x}$ is the sample mean. 
The bottom left panel of Fig. \ref{fig:filament-galaxy} shows the skewness of the distribution in each $D_{\rm fil}$ bin. The error bars are the 95\% confidence level obtained by bootstrap resampling 10000 times. 
The dashed horizontal line in pink colour indicates the zero skewness as a reference. 
We confirm that the filament-galaxy angle distribution is positively skewed within $D_{\rm fil} \lesssim 11\,\rm Mpc$. The skewness is the highest in the bin closest to the filament ($D_{\rm fil}\lesssim6\,\rm Mpc$). At large distances beyond $D_{\rm fil}\gtrsim 11\,\rm Mpc$, the skewness fluctuates around zero.

We perform the Kolmogorov-Smirnov (KS) test to examine the statistical significance of our findings. 
First, we use the one-sample KS test to compare the filament-galaxy angle distribution in each distance bin with the uniform distribution. 
In the bottom right panel of Fig. \ref{fig:filament-galaxy}, we show the p-value of the KS test in each bin with a solid horizontal line spanning the corresponding $D_{\rm fil}$ range. The dashed horizontal line in pink colour is a p-value of 0.05 for reference. 
The p-value is 0.00073 in $D_{\rm fil}\leq6.36\,\rm Mpc$ and 0.0145 in $6.36\,\rm Mpc<D_{\rm fil}\leq10.96\,\rm Mpc$. 
In both cases, we reject the null hypothesis that the filament-galaxy angle distribution is uniform with a confidence level of $\gtrsim99$ per cent. 
In larger $D_{\rm fil}$ ranges, all p-values are greater than 0.05. We cannot reject the hypothesis that these distributions are uniform. 

We also perform the two-sample KS test comparing the distributions within the first bin ($D_{\rm fil}\leq6.36\,\rm Mpc$) and the rest of the sample ($6.36\,\rm Mpc<D_{\rm fil}<80\,\rm Mpc$). 
The p-value of this test is 0.002 (the red star symbol in the first $D_{\rm fil}$ bin in Fig. \ref{fig:filament-galaxy}). We therefore reject the null hypothesis that the distribution of filament-galaxy angle in the $D_{\rm fil}\leq6.36\,\rm Mpc$ bin is drawn from the same underlying distribution as the rest of the sample with a confidence level of $>99$ per cent.

Next, we investigate whether the systematic galaxy-filament alignment among galaxies close to filaments depends on the optical ellipticity of the galaxy, $|\epsilon_{\rm opt}|$ (see Equation \ref{eq:ellip_opt} for the definition of $\epsilon_{\rm opt}$).
The top panel of Fig. \ref{fig:filament-galaxy_ellip_incl} shows the skewness of the filament-galaxy angle distribution at different $D_{\rm fil}$ bins with the GMRG sample divided into two subsamples depending on the optical ellipticity: $|\epsilon_{\rm opt}|\leq0.1$ (black line) and $|\epsilon_{\rm opt}|>0.1$ (yellow line). Only the inner $D_{\rm fil}<25\,\rm Mpc$ range is shown for visualization purposes. 
The blue colour line shows the result from the entire GMRG sample regardless of their ellipticities (i.e., the same line shown in the bottom left panel of Fig. \ref{fig:filament-galaxy}), for reference. 
We find the increase in the skewness at the innermost bin ($D_{\rm fil} \leq 6.36\,\rm Mpc$) compared to the other bins is higher for the sample with higher optical ellipticities. 
The p-value of the one-sample KS test between the uniform distribution and the filament-galaxy angle distribution at $D_{\rm fil}\leq6.36\,\rm Mpc$ and $|\epsilon_{\rm opt}|>0.1$ is $2.5\times10^{-5}$.
In contrast, galaxies with $|\epsilon_{\rm opt}|\leq0.1$ do not reveal any sign of systematic alignment with their closest filament in the skewness at all distances. The p-value of the one-sample KS test, in this case between the uniform distribution and the filament-galaxy angle distribution at $D_{\rm fil}\leq6.36\,\rm Mpc$ and $|\epsilon_{\rm opt}|\leq0.1$, is 0.43.

Finally, we test whether the observed galaxy-filament alignment depends on the inclination of cosmic filaments ($i_{\rm fil}$) with respect to the sky plane. Higher $i_{\rm fil}$ means that the filaments are well aligned with the radial sightline, and lower $i_{\rm fil}$ means that the filaments are parallel to the sky plane. 
The bottom panel of Fig. \ref{fig:filament-galaxy_ellip_incl} shows the skewness of the filament-galaxy angle distribution at different $D_{\rm fil}$ bins, with samples divided into $i_{\rm fil}\leq45^{\circ}$ (black line) and $i_{\rm fil}>45^{\circ}$ (yellow line). 
In the innermost bin ($D_{\rm fil} \leq 6.36\,\rm Mpc$), we find that the sample with $i_{\rm fil}\leq45^{\circ}$ shows higher skewness than its low inclination counterpart. 
The one-sample KS test rejects the null hypothesis that the filament-galaxy angle distribution is uniform (p-value of $4.4\times10^{-4}$) in this sample. 
For the sample with $i_{\rm fil}>45^{\circ}$, the skewness is close to zero in all $D_{\rm fil}$ ranges. The p-value of the one-sample KS test between the uniform distribution and the filament-galaxy angle distribution at $D_{\rm fil}\leq6.36\,\rm Mpc$ and $i_{\rm fil}>45^{\circ}$ is 0.6. 

In brief summary, we find strong evidence that the galaxies in the cosmic filament environment ($D_{\rm fil}\lesssim11\,\rm Mpc$) have their major axis aligned with the orientation of the filaments.
The alignment is stronger among galaxies with higher optical ellipticity and filaments with smaller inclinations (i.e., more parallel to the sky plane). 
In Section \ref{sec:mechanism_filament}, we will discuss the $\epsilon_{\rm opt}$ and $i_{\rm fil}$ dependencies based on (i) the projection effect; (ii) the intrinsic shape of galaxies; and (iii) observational uncertainty in the 3D orientation of filaments.

\subsection{Alignment between optical galaxy and radio jet}\label{sec:jet-galaxy}

\begin{figure*}
    \centering
    \includegraphics[width=0.95\textwidth]{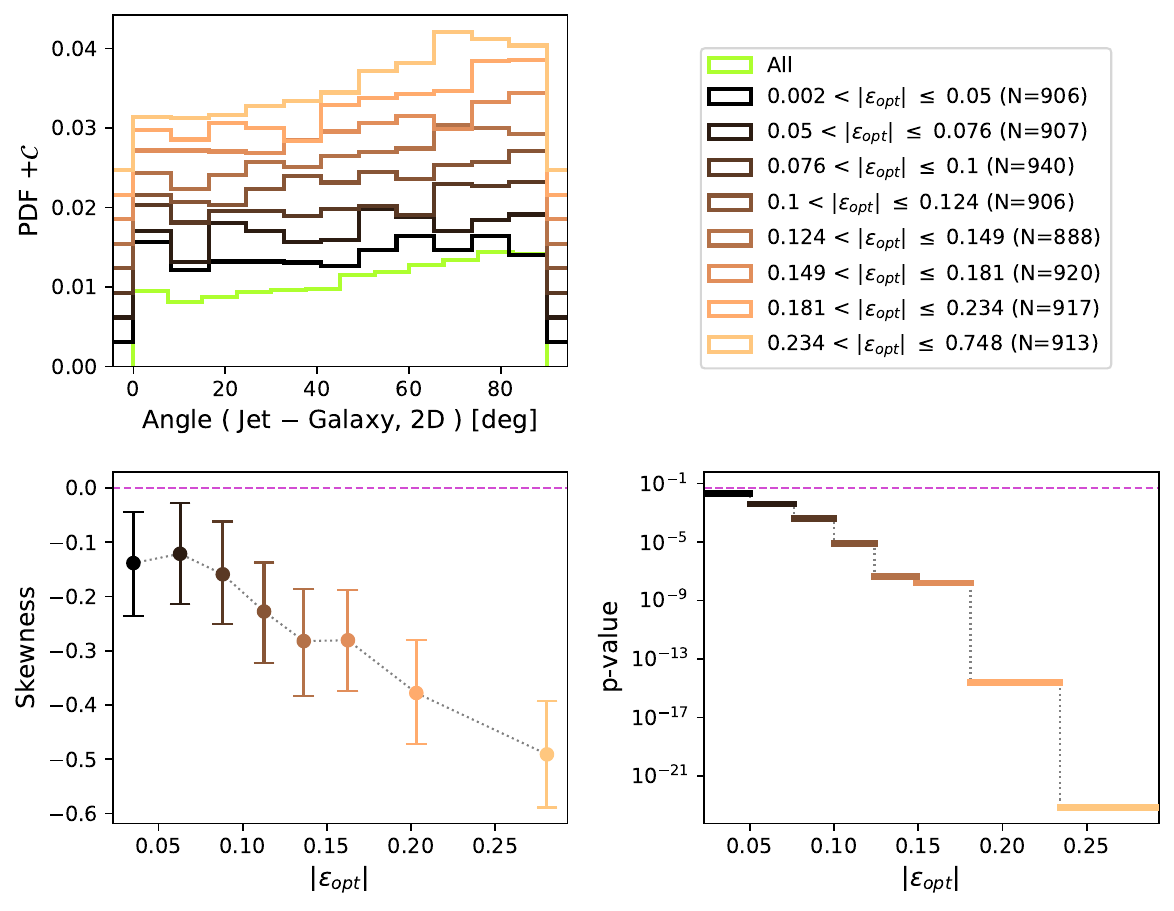}
    \caption{
    The distribution of the angle between the galaxy optical major axis and the radio jet orientation as a function of the optical ellipticity. The EJ sample is used for this analysis. The panels are in the same format as Fig. \ref{fig:filament-galaxy}. 
    }  
    \label{fig:jet-galaxy_ellipticity}
\end{figure*}

\begin{figure*}
    \centering
    \includegraphics[width=0.95\textwidth]{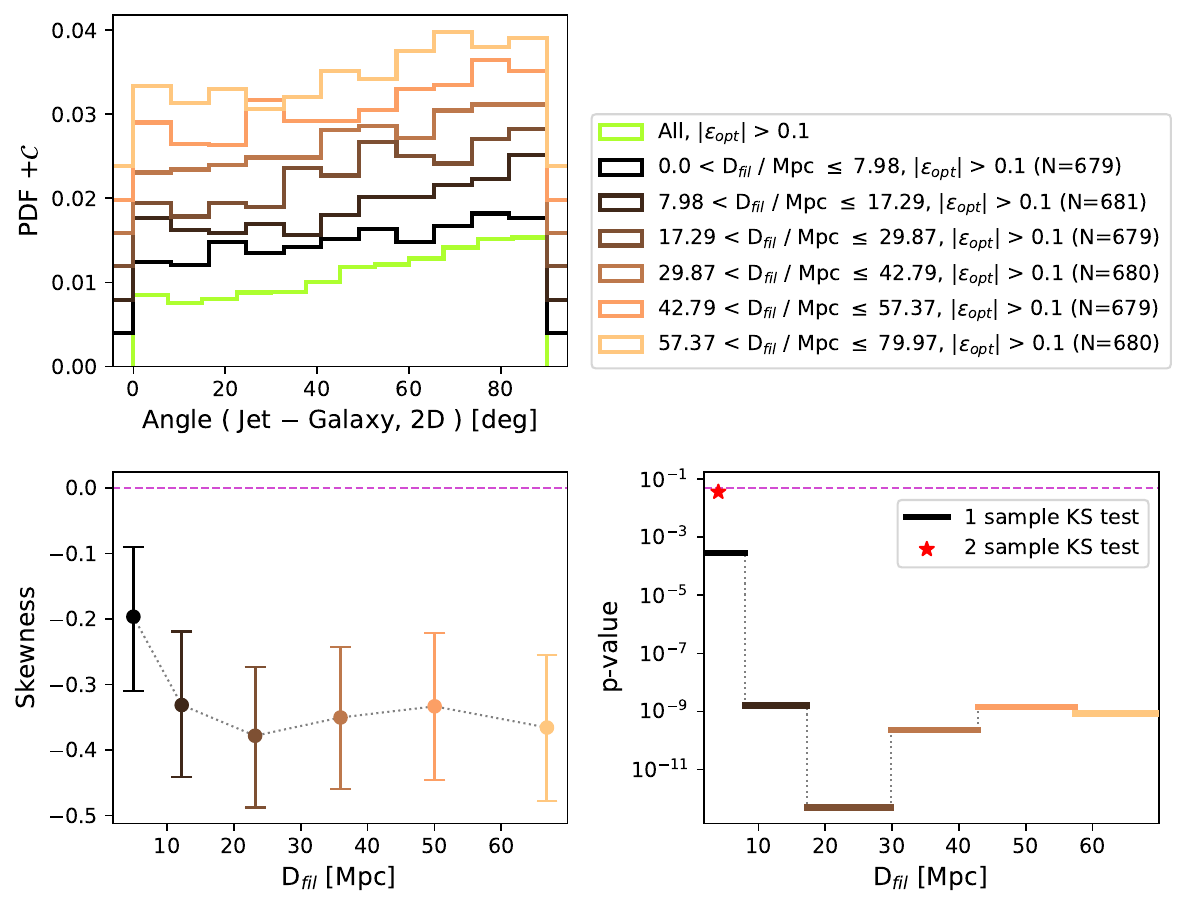}
    \caption{
    The distribution of the angle between the galaxy optical major axis and the radio jet orientation as a function of the distance to the filament. The EJ sample is used for this analysis. The panels are in the same format as Fig. \ref{fig:filament-galaxy}. 
    }  
    \label{fig:jet-galaxy}
\end{figure*}

In this section, we investigate the angle between the radio jets in the EJ sample and the major axis of their optical host galaxies.
Our motivation to do so is to understand whether the optical galaxy-cosmic filament alignment within the filaments we find in the previous section is due to a direct connection between the evolution of a galaxy’s stellar component and cosmic filaments or if it is a secondary correlation that arises from interactions between radio jets and the intergalactic medium (IGM), along with the relationship between jet and galaxy orientations. 
Unlike the previous analysis comparing the orientations of filaments and galaxies, we do not implement the parallel transport method in this case, as the angular separations between the cross-matched radio and optical pairs are co-spatial.

First, we examine the effect of optical ellipticity on the jet-galaxy alignment. 
The upper left panel of Fig. \ref{fig:jet-galaxy_ellipticity} shows the histograms of the jet-galaxy angle in different optical ellipticity bins. 
The green line represents the entire EJ sample within $80\,\rm Mpc$ of the closest filaments. 
The distribution peaks at $90^{\circ}$, indicating that the radio jets tend to align perpendicularly to the galactic major axis in general. 
Histograms in other colours that represent different optical ellipticity bins are shifted along the $y$-axis for better visibility. 
For galaxies in the lowest ellipticity range, i.e., $|\epsilon_{\rm opt}|<0.05$, the angle between the radio jet and the optical major axis is nearly randomly distributed between 0 and $90^{\circ}$ (see the histogram in the black line, which is close to the uniform distribution). 
With increasing ellipticity, the histograms show a stronger peak at $90^{\circ}$. 
The skewness of the jet-galaxy angle distribution is presented in the bottom left panel of Fig. \ref{fig:jet-galaxy_ellipticity}. This result further confirms the trend: with increasing $|\epsilon_{\rm opt}|$, the distribution is more negatively skewed, i.e., more strongly skewed toward $90^{\circ}$.
This result aligns with previous findings of \citet{Zheng_2024}, where the authors find stronger alignment between radio jets and optical minor axis among galaxies with smaller optical minor-to-major axes ratio (e.g., see Fig. 10 of their paper). In their paper, the authors attribute this trend to the projection effect of jets largely aligned with the minor axis of an oblate galaxy.

Similarly, the results of the one-sample KS test shown in the bottom right panel of Fig. \ref{fig:jet-galaxy_ellipticity} confirm that the p-values of the test decrease significantly with increasing $|\epsilon_{\rm opt}|$. 
This means that we can reject the null hypothesis that the jet-galaxy angles are drawn from the uniform distribution with an increasingly higher confidence level as the optical ellipticity increases. 
Nonetheless, p-values are always lower than 0.05 in all $|\epsilon_{\rm opt}|$ ranges; the jet-galaxy angle distribution is always nonuniform at a confidence level of 95\%.

Based on the above results, we select a subset of EJ sample galaxies with $|\epsilon_{\rm opt}|>0.1$ as a subsample with substantial bias towards having a jet perpendicular to the galaxy major axis. 
For the remainder of this section, we investigate whether cosmic filaments have an effect on this bias. 

In Fig. \ref{fig:jet-galaxy}, we show the jet-galaxy angle distribution in various $D_{\rm fil}$ bins in the same format as Figs. \ref{fig:filament-galaxy} and \ref{fig:jet-galaxy_ellipticity}. 
As can be noticed from histograms in the top left panel and the skewness distribution in the bottom left panel, the distribution of jet-galaxy angle is less strongly peaked at $90^{\circ}$ in the bin closest to filaments ($D_{\rm fil}\lesssim8\,\rm Mpc$) compared to the other larger distance ranges. The skewness of the distribution is the lowest in this $D_{\rm fil}$ range. 
The skewness remains nearly constant in all other $D_{\rm fil}$ ranges beyond $8\,\rm Mpc$.

An alternative way to quantify the jet-galaxy angle distribution is to calculate the fraction of galaxies with `misaligned' jets. Supposing jets are intrinsically aligned with the optical minor axis, we consider them to be misaligned if the position angle is more than $30^{\circ}$ away from the optical minor axis ($<60^{\circ}$ in the histogram since the angle in Fig. \ref{fig:jet-galaxy} is measured from the optical major axis). We find $\approx60.2^{+3.5}_{-3.7}\%$ of the EJ galaxies have misaligned jets at $D_{\rm fil} \lesssim8\,\rm Mpc$. The fraction decreases to $\approx56.2^{+1.7}_{-1.6}\%$ at larger distance. Potentially, the fraction of misaligned jets will be larger among galaxies even closer to filament spines than $8\,\rm Mpc$. However, we could not confirm the statistical significance due to the limited sample size.

The bottom right panel of Fig. \ref{fig:jet-galaxy} shows the outcomes of the KS tests. 
The p-values of the one-sample KS tests (solid horizontal lines) are always lower than 0.05 (dashed horizontal line). This means that, regardless of the distance to the closest filament, we reject the null hypothesis that the jet-galaxy angles are drawn from the uniform distribution with a 95\% confidence level. 
This is expected since, as mentioned above, we have purposely selected a subpopulation of the EJ sample that is supposed to show a clear preference for the jets being perpendicular to the galaxy major axis. 
The p-value of the two-sample KS test between the jet-galaxy angle distribution of galaxies in $D_{\rm fil}<8\,\rm Mpc$ and $>8\,\rm Mpc$ is 0.036 (red star symbol). 
We reject the null hypothesis that the jet-galaxy angles in two $D_{\rm fil}$ ranges are drawn from the same distribution with a 95\% confidence level. 

Based on the results presented in this section, we conclude that galaxies in the cosmic filament environment ($D_{\rm fil}\lesssim 8\,\rm Mpc$) have radio jets that are more randomly oriented with respect to the galaxy optical major axis than galaxies at larger distances to filaments. Still, the jet direction is not entirely random.

\section{Discussion}

\subsection{Physical picture}\label{sec:mechanism}

Probing alignment between radio jets, optical galaxies, and the large-scale structure provides interesting insight into galaxy evolution. 
The direction of radio jets is related to the accretion mode of the central black hole as will be explained in detail later in this section. 
The observed shape of optical galaxies is closely related to their stellar kinematics shaped by the underlying gravitational potential of their halo, as well as their past star formation and merger events. 
Cosmic filaments shape the large-scale matter flow pattern surrounding galaxies. 
In this section, we summarise previous theoretical studies that explain the alignment/misalignment of (i) galaxies with respect to cosmic filaments and (ii) AGN jets with respect to galaxies. We then discuss how the results presented in Section \ref{sec:result} fit into the physical picture.

\subsubsection{Mechanisms for filament-optical galaxy alignment}\label{sec:mechanism_filament}

In numerical simulations, dark matter halos are found to be more intrinsically elongated and prolate with increasing mass (\citealt{Bett_2007, Tenneti_2014}). This suggests that the build-up of these massive halos is affected by directional, filamentary accretion of mass (\citealt{Allgood_2006, Vera-Ciro_2011}).  
Although it has been reported that massive halos experience a transition of accretion pattern throughout their evolution from directional to isotropic accretion (\citealt{Allgood_2006}), the directional accretion is still significant in cosmic filament environments where halos travel along the filaments towards denser nodes and collide with each other. 
In this case, the direction of mergers, accretion, and fly-bys that halos experience throughout their evolution is predominantly aligned with the orientation of the cosmic filament. 
As a result, the major axis of halos tends to align with the cosmic filaments in which they are embedded, and the significance of the alignment increases with halo mass (\citealt{Lee_2008, Zhang_2009,Libeskind_2014, Kang_2015,Morinaga_2020}). 

The alignment between galaxies and cosmic filaments should be interpreted in a similar context. 
In the stellar mass range probed in this study, $\log M_{\rm *}/\rm M_{\odot}>11$, galaxies and their host halo grow by numerous mergers, and there is a high fraction of dispersion-dominated slow-rotating galaxies (\citealt{Emsellem_2011, Veale_2017, vandesande_2017}). 
These systems are flattened by the anisotropic velocity dispersion of the stars rather than by their rotational motion (\citealt{Binney_1978, Binney_2008}). 
For that reason, the optical position angle of the galaxies we use in this study is closely related to the shape of the gravitational potential and the orbital angular momentum of previously accreted subhalos/galaxies. 
In the following paragraphs, we discuss how our results support the scenario that, for galaxies that reside within the cosmic filament environment, mergers have occurred predominantly along the direction of the filaments.

In Section \ref{sec:filament-galaxy}, we find that the optical major axis of galaxies within $11\,\rm Mpc$ of cosmic filaments is systematically oriented parallel to the closest cosmic filament. 
If this galaxy-filament alignment is caused by physical processes, one can expect the following tendency:
(i) The alignment is strong within the cosmic filaments and nonexistent outside the boundary of filaments; (ii) The projection effect affects the significance of the alignment signal observed on the 2D sky plane. 
Indeed, we confirm that both trends are present in our sample. 
First, we find the galaxy-filament alignment is the strongest among galaxies closer than $6\,\rm Mpc$ from the filaments (see Fig. \ref{fig:filament-galaxy}).
There is a weaker sign of alignment between 6 and $11\,\rm Mpc$ of the filaments. Beyond $11\,\rm Mpc$, galaxies are randomly oriented with respect to the matched filament because there is no preferred direction of mergers set by nearby filaments. 
From this result, we can infer that the characteristic radius of cosmic filaments, where directional mergers and accretion take place, is $\lesssim 6-11\,\rm Mpc$. This result is in broad agreement with previous investigations on the characteristic radius of filaments (e.g., \citealt{Bond_2010, Cautun_2014, Bonjean_2020, Galarraga-Espinosa_2020}), though the exact scale is subject to the definition of the boundary (\citealt{Wang_2024}).

Second, we test the projection effect by comparing the significance of the observed galaxy-filament alignment depending on the inclination of the filaments and the galaxy ellipticity (see Fig. \ref{fig:filament-galaxy_ellip_incl}). 
The idea is that filaments with close-to-zero inclination, i.e., filaments oriented along the plane of the sky, will reveal an alignment as strong as the 3D alignment intrinsically arising from physical processes, whereas filaments with their inclination close to $90^{\circ}$ will not reveal any sign of alignment since the intrinsic 3D axis of alignment is parallel to the line of sight. 
Similarly, edge-on galaxies will show a stronger alignment on the sky plane than face-on galaxies, as their intrinsic major axis is parallel to the plane of the sky. 
Although we do not have a direct tracer for the inclination of optical galaxies, we take advantage of the fact that at a given intrinsic ellipticity, higher projected ellipticity corresponds to higher galaxy inclination, i.e., the edge-on view. 
We find a stronger galaxy-filament alignment in the cosmic filament environment with increasing optical ellipticity and decreasing filament inclination.
This further supports our conclusion that the galaxy-filament alignment is of physical origin.

Note that galaxies with higher optical ellipticity can also be the ones with higher intrinsic ellipticity. 
In this case, the stronger major-axis alignment with filaments among galaxies with higher optical ellipticity can be interpreted such that more intrinsically elongated galaxies have a stronger alignment with cosmic filaments. The higher intrinsic ellipticity might indicate a more coherent direction of mergers throughout the assembly of galaxies. However, it is beyond the scope of this paper to separate how much of the optical ellipticity dependency of the galaxy-filament alignment comes from the projection effect and the change in intrinsic ellipticity.

The stronger galaxy-filament alignment we find among galaxies matched with filaments with lower inclinations could be in part related to how well the orientation of filaments in three dimensions is defined observationally. 
It is more likely that filaments with smaller inclinations will have their orientation more accurately defined, as (i) we are more precise in determining the RA and Dec. of filaments compared to the spectroscopic redshifts and (ii) there is inevitable ambiguity in determining the true distance due to the peculiar velocities of galaxies.

Although we have confirmed a statistically significant alignment signal between galaxies and filaments, it is worth pointing out that there is a high level of randomness in the galaxy-filament angle distribution.
This indicates that processes that randomize the filament-galaxy angle are common even in the cosmic filament environment.
One example of such processes is mergers that do not occur along the direction of filaments. 
If there was a relatively recent merger event that produced stellar shells and streams, it can affect the optical position angle measurements. 
The tidal debris can live for several billion years before it disperses and settles into the gravitational potential of the central galaxies (\citealt{Quinn_1984, Hendel_2015, Pop_2018}).

\subsubsection{Mechanisms for jet-optical galaxy alignment}\label{sec:mechanism_jet}

All extended radio jets in our sample are produced by the SMBH at the centre of galaxies. 
The direction of radio jets powered by the accreting material is expected to align with the spin axes of the black hole which is affected by the angular momentum of gas in the inner accretion disc (\citealt{Bardeen_1975}). 
If gas accreting onto the black hole has a comparable angular momentum with galactic gas at larger scales, the direction of jets would be aligned with the angular momentum of the host galaxy (``secular accretion''). 
However, the angular momentum of gas related to individual accretion events is not always equivalent to that of the larger scale gas motion (\citealt{King_2007}). 
Mergers and secular gas disc instabilities can generate large-scale perturbations that lead to chaotic accretion of gas onto black holes (\citealt{Hobbs_2011,Hopkins_2012}).

In Section \ref{sec:jet-galaxy}, we find radio jets of our sample galaxies are overall perpendicular to the optical major axis (see Fig. \ref{fig:jet-galaxy}). This result suggests that secular accretion is in action. This is in line with previous observational studies by \citet{Battye_2009}; \citet{Zheng_2024}; \citet{Gil_2024}.
However, we also confirm that there is a notable level of randomness in the jet-galaxy angle distribution. 
Specifically, the randomness is higher among galaxies closer than $8\,\rm Mpc$ to the nearest cosmic filaments. 
As discussed in the previous section, galaxies in our sample are likely to have experienced numerous mergers while travelling along cosmic filaments.
The angular momentum of nuclear gas accreting onto the black hole, as well as the black hole spin itself, can be significantly disturbed by galaxy-galaxy interactions. 
This leads to more chaotic accretion that produces randomly oriented jets with respect to the galactic stellar and gas distribution.

\subsection{Implication}\label{sec:implication}

\subsubsection{Absence of large-scale alignment of radio jets}

In recent years, there has been an ongoing search for coherent orientation of radio jets over a large area of the sky to test the influence of large-scale cosmic environment on the evolution of galaxies and accretion of gas onto the SMBH. However, the results are contentious (\citealt{Taylor_2016, Contigiani_2017, Osinga_2020, Panwar_2020, Mandarakas_2021, Simonte_2023}).

We have shown in this paper that (i) the major axis of galaxies tends to align with the closest cosmic filament if a galaxy is closer than $\lesssim11\,\rm Mpc$ from the filaments and (ii) radio jets are generally perpendicular to the major axis of the galaxy. Therefore, any putative large-scale alignment among observed radio jets can be attributed to the combined effect of these two.
However, when more carefully considering the environment galaxies reside in, we find that the preference towards jets aligning perpendicular to the galaxy's major axis weakens in environments close to filaments ($\lesssim8\,\rm Mpc$). The physical mechanisms behind this are discussed in Section \ref{sec:mechanism}. 
Combining these results together, we argue that it is hard to expect a strong large-scale alignment of radio jets caused by cosmic filaments.

\subsubsection{Intrinsic alignment of galaxies}

Weak lensing quantifies the correlation between the distortion of galaxy shapes on the sky plane, i.e. cosmic shear, to infer the mass distribution along the sightlines. 
It is one of the key science projects for ongoing and upcoming large observing programs, such as the Dark Energy Survey (\citealt{DES_2021}), the Vera Rubin Observatory (\citealt{Ivezic_2019}), and \textit{Euclid} (\citealt{Laureijs_2011,Amendola_2018, Euclid_2024}). 
Under the assumption that the Universe is homogeneous and isotropic, the orientation of galaxies is expected to be random before the light is lensed.

However, a non-negligible correlation between galaxy ellipticities, i.e., the intrinsic alignment of galaxies, indeed affects weak lensing cosmic shear measurements. 
Previous studies have shown that the intrinsic alignment between galaxies is stronger for bright galaxies (\citealt{Hirata_2007, Joachimi_2011, Singh_2015}). 
In Section \ref{sec:filament-galaxy}, we show that radio galaxies with stellar mass $>10^{11}\,\rm M_{\odot}$ tend to align with the closest cosmic filament. This is direct evidence that cosmic filaments are at least in part responsible for imposing a locally preferred orientation on galaxies. 
Though we do not aim to quantify the pairwise alignment between galaxies in this work, our results highlight the importance of accounting for intrinsic alignment to accurately interpret cosmological weak lensing surveys (\citealt{Heavens_2000, Croft_2000}).

Furthermore, many studies have shown that feedback from AGN, and specifically mechanical feedback from jets at $z<1$, can disrupt the mass distribution within the dark-matter halo \citep[e.g][]{Peirani2017,Chisari2018, Foreman2020}. Our work suggests that the impact of jets, in galaxies that are most aligned with the large-scale filamentary structures, will also preferentially deposit energy along the minor axis of the dark matter halo that is tidally aligned with the large-scale structure. This would suggest that accurately incorporating AGN feedback effects on small scales in cosmological analyses will depend strongly on whether the galaxies are lying within a filament or not.

\subsubsection{Azimuthal anisotropy within the boundary of galactic halos}

Many spectral line studies on diffuse gas suggest that the observed distribution of the multiphase circumgalactic medium (CGM) around galaxies is not azimuthally isotropic (e.g., \citealt{Bordoloi_2011, Bouche_2012, Kacprzak_2012, Kacprzak_2015, Huang_2016}). 
Specifically, some studies along this line focus on massive galaxy populations (\citealt{Zhang_2018, Zhang_2022}) and suggest that the inner CGM of these systems, especially along the direction of the galaxy minor axis, is ionized by radiation from AGN activities. 
Such asymmetries in the CGM properties are predicted by cosmological hydrodynamic simulations (e.g., \citealt{Peroux_2020, Truong_2021, Nelson_2021, Pillepich_2021, Ramesh_2023}) and are often attributed to the result of bipolar outflow along the minor axis of galaxies, either or both from the AGN and stellar feedback, and cold inflow along the major axis of galaxies.

According to the analytic calculation by \citet{RawlingsJarvis2004}, the amount of energy injected by radio mode AGN feedback is powerful enough to gravitationally unbind the CGM of the AGN host galaxy as well as its neighbouring satellite galaxies.
Based on our finding that radio jets tend to align with the optical minor axis of galaxies in a statistically large sample, we support the scenario that radio-mode AGN feedback is one of the mechanisms shaping the anisotropy in the CGM around massive radio galaxies. 
However, we suggest that such an AGN contribution could depend on the cosmic environment since we observe the direction of radio jets becomes more random with respect to the orientation of the optical host galaxy the closer it gets to cosmic filaments.
On the other hand, we find that the optical major axis of galaxies is preferentially aligned with the direction of the cosmic filaments they reside within. This trend suggests that the accretion of matter onto these galaxies brought by cosmic filaments is directed along the galactic major axis and supports the scenario that the cold accretion of the IGM could be responsible for the anisotropy of the CGM.

Azimuthal anisotropy is found not only in CGM properties but also in satellite galaxy populations in galaxy clusters. 
A number of recent studies demonstrated that there is an excess of star-forming, blue satellite galaxy population along the minor axis of the brightest cluster galaxies (BCGs) compared to the major axis (\citealt{Huang_2016, Martin_Navarro_2021, Stott_2022, Ando_2023, Karp_2023, Stephenson_2025}). 
Several physical mechanisms have been proposed to explain this trend. First, \citet{Martin_Navarro_2021} suggest that satellite galaxies could experience weaker ram pressure along the minor axis where AGN outflows take place. As mentioned in the previous paragraph, our results showing the alignment of radio jets with galaxy minor axes support this picture that AGN feedback, at least to some degree, contributes to shaping the anisotropic CGM structure that could lead to the differential ram pressure stripping along different axes.
On the other hand, \citet{Stott_2022} suggest an alternative scenario in which the elongated shape of the clusters itself causes the differential ram pressure stripping towards the major and minor axes. 
Furthermore, recent work by \citet{Stephenson_2025} shows that the excessive population of quenched galaxies along the BCG major axis extends out to several virial radii of the clusters in their sample. They explain this to be a combined outcome of (i) satellite galaxies undergoing quenching in cosmic filaments before falling into clusters and (ii) the BCG major axis aligning with the large-scale structure (see also theoretical works by \citealt{Karp_2023, Zakharova_2025}). 
Though in this work we do not address the former, which is often referred to as pre-processing, there is ample evidence that cosmic filaments do affect the star formation of galaxies (\citealt{Sarron_2019, Hoosain_2024}) and a significant fraction of satellite systems in clusters assemble through this channel (\citealt{Jung_2018, Han_2018, Kuchner_2022}). 
As for the second point, we have shown that the optical major axis of massive galaxies in cosmic filament environments tends to align with the direction of the nearest filament. Therefore, we support the proposition that the azimuthal segregation of satellite galaxies based on their star-formation status can be a natural outcome of the hierarchical formation of the large-scale structure of the Universe.

\section{Summary}\label{sec:summary}

In this paper, we analyse the position angles of the optical and radio major axes and the orientation of filaments closest to galaxies. We use DESI Legacy Imaging Surveys for the optical data, LoTSS DR2 radio-optical cross-matching catalogue for the radio data, and the DisPerSE filament finder algorithm on SDSS DR12 galaxy distribution for the filament catalogue. 
All galaxies used in this study are radio galaxies with stellar mass above $M_{*}>10^{11}\,\rm M_{\odot}$. For the analysis of radio jet position angles, we select a subsample with bright extended radio sources. 
The distance between galaxies and filaments is measured in 3D.
The position angles are compared in the projected sky plane, and we apply the parallel transport method wherever appropriate.

Here are the main findings of this paper.
\begin{enumerate}[wide = 0pt]
    \item In the cosmic filament environment, more specifically, within $\lesssim11\,\rm Mpc$ of filaments, a galaxy's major axis tends to align with the orientation of the closest filament. This galaxy-filament alignment is more robust in the regions closer to the filament spine and is affected by the projection effect. 

    \item Radio jets are, in general, aligned perpendicularly to the major axis of their host galaxy. However, within $\lesssim8\,\rm Mpc$ from cosmic filaments, jets are more randomly oriented with respect to the optical major axis.
\end{enumerate}

These results have three major implications. 
First, massive galaxies in the cosmic filament environment grow by directional merger along the filaments they reside in. We expect the impact parameter of cosmic filaments in this regard to be $\lesssim 11\,\rm Mpc$.
Mergers along filaments lead to the large-scale elongation of the dark matter halo and stellar distribution, resulting in the alignment of the optical major axis and the filament orientation. The spin axis of the SMBH is more easily affected by individual merger events, and the chaotic accretion onto the black hole produces randomly directed jets. 

Second, cosmic filaments could be responsible for the large-scale alignment of radio jets if there is any.
The search for the alignment among adjacent jets has been a topic of interest with the rise of large-area radio surveys as a test for the connection between cosmic large-scale environments and the growth of SMBH. 
However, we demonstrate that the effect of cosmic filaments on the SMBH feedback manifests itself in the more randomised angle between jets and their host galaxies. 
Therefore, we expect the alignment signal to be extremely weak due to environmental effects on jet orientation. 

Finally, the alignment of radio jets with the optical minor axis suggests that radio-mode AGN feedback could be one of the mechanisms driving the observed azimuthal anisotropy distribution of the CGM. Meanwhile, the alignment of the optical major axis with the filament orientation is related to the hierarchical assembly of clusters within the large-scale structure. Both mechanisms have been suggested to explain the observed azimuthal segregation of satellite galaxies in galaxy clusters based on their star formation status, though, they are likely to work at different radii.

\section*{Acknowledgements}
SLJ, MJJ, MNT, and TY acknowledge the support of a UKRI Frontiers Research Grant [EP/X026639/1], which was selected by the European Research Council, and the STFC consolidated grants [ST/S000488/1] and [ST/W000903/1]. IHW, MJJ, CLH, and MNT also acknowledge support from the Oxford Hintze Centre for Astrophysical Surveys which is funded through generous support from the Hintze Family Charitable Foundation.  

Our analysis was performed using the Python programming language (Python Software Foundation, https://www.python.org). The following packages were used throughout the analysis: numpy (\citealt{Harris_2020}), SciPy (\citealt{Virtanen_2020}), matplotlib (\citealt{Hunter_2007}). This work also made use of Astropy:\footnote{http://www.astropy.org} a community-developed core Python package and an ecosystem of tools and resources for astronomy \citep{astropy:2013, astropy:2018, astropy:2022}.

LoTSS DR 2 data products were provided by the LOFAR Surveys Key Science project (LSKSP; https://lofar-surveys.org/) and were derived from observations with the International LOFAR Telescope (ILT). LOFAR (\citealt{vanHaarlem_2013}) is the Low Frequency Array designed and constructed by ASTRON. It has observing, data processing, and data storage facilities in several countries, which are owned by various parties (each with their own funding sources), and which are collectively operated by the ILT foundation under a joint scientific policy. The efforts of the LSKSP have benefited from funding from the European Research Council, NOVA, NWO, CNRS-INSU, the SURF Co-operative, the UK Science and Technology Funding Council and the Jülich Supercomputing Centre.

The Legacy Surveys consist of three individual and complementary projects: the Dark Energy Camera Legacy Survey (DECaLS; Proposal ID \#2014B-0404; PIs: David Schlegel and Arjun Dey), the Beijing-Arizona Sky Survey (BASS; NOAO Prop. ID \#2015A-0801; PIs: Zhou Xu and Xiaohui Fan), and the Mayall z-band Legacy Survey (MzLS; Prop. ID \#2016A-0453; PI: Arjun Dey). DECaLS, BASS and MzLS together include data obtained, respectively, at the Blanco telescope, Cerro Tololo Inter-American Observatory, NSF’s NOIRLab; the Bok telescope, Steward Observatory, University of Arizona; and the Mayall telescope, Kitt Peak National Observatory, NOIRLab. Pipeline processing and analyses of the data were supported by NOIRLab and the Lawrence Berkeley National Laboratory (LBNL). The Legacy Surveys project is honored to be permitted to conduct astronomical research on Iolkam Du’ag (Kitt Peak), a mountain with particular significance to the Tohono O’odham Nation.

NOIRLab is operated by the Association of Universities for Research in Astronomy (AURA) under a cooperative agreement with the National Science Foundation. LBNL is managed by the Regents of the University of California under contract to the U.S. Department of Energy.

This project used data obtained with the Dark Energy Camera (DECam), which was constructed by the Dark Energy Survey (DES) collaboration. Funding for the DES Projects has been provided by the U.S. Department of Energy, the U.S. National Science Foundation, the Ministry of Science and Education of Spain, the Science and Technology Facilities Council of the United Kingdom, the Higher Education Funding Council for England, the National Center for Supercomputing Applications at the University of Illinois at Urbana-Champaign, the Kavli Institute of Cosmological Physics at the University of Chicago, Center for Cosmology and Astro-Particle Physics at the Ohio State University, the Mitchell Institute for Fundamental Physics and Astronomy at Texas A\&M University, Financiadora de Estudos e Projetos, Fundacao Carlos Chagas Filho de Amparo, Financiadora de Estudos e Projetos, Fundacao Carlos Chagas Filho de Amparo a Pesquisa do Estado do Rio de Janeiro, Conselho Nacional de Desenvolvimento Cientifico e Tecnologico and the Ministerio da Ciencia, Tecnologia e Inovacao, the Deutsche Forschungsgemeinschaft and the Collaborating Institutions in the Dark Energy Survey. The Collaborating Institutions are Argonne National Laboratory, the University of California at Santa Cruz, the University of Cambridge, Centro de Investigaciones Energeticas, Medioambientales y Tecnologicas-Madrid, the University of Chicago, University College London, the DES-Brazil Consortium, the University of Edinburgh, the Eidgenossische Technische Hochschule (ETH) Zurich, Fermi National Accelerator Laboratory, the University of Illinois at Urbana-Champaign, the Institut de Ciencies de l’Espai (IEEC/CSIC), the Institut de Fisica d’Altes Energies, Lawrence Berkeley National Laboratory, the Ludwig Maximilians Universitat Munchen and the associated Excellence Cluster Universe, the University of Michigan, NSF’s NOIRLab, the University of Nottingham, the Ohio State University, the University of Pennsylvania, the University of Portsmouth, SLAC National Accelerator Laboratory, Stanford University, the University of Sussex, and Texas A\&M University.

BASS is a key project of the Telescope Access Program (TAP), which has been funded by the National Astronomical Observatories of China, the Chinese Academy of Sciences (the Strategic Priority Research Program ``The Emergence of Cosmological Structures'' Grant \# XDB09000000), and the Special Fund for Astronomy from the Ministry of Finance. The BASS is also supported by the External Cooperation Program of Chinese Academy of Sciences (Grant \# 114A11KYSB20160057), and Chinese National Natural Science Foundation (Grant \# 12120101003, \# 11433005).

The Legacy Survey team makes use of data products from the Near-Earth Object Wide-field Infrared Survey Explorer (NEOWISE), which is a project of the Jet Propulsion Laboratory/California Institute of Technology. NEOWISE is funded by the National Aeronautics and Space Administration.

The Legacy Surveys imaging of the DESI footprint is supported by the Director, Office of Science, Office of High Energy Physics of the U.S. Department of Energy under Contract No. DE-AC02-05CH1123, by the National Energy Research Scientific Computing Center, a DOE Office of Science User Facility under the same contract; and by the U.S. National Science Foundation, Division of Astronomical Sciences under Contract No. AST-0950945 to NOAO.

This work makes use of the service L3S/COSFIL, the Large Scale Structure Services developed by OSUPS, OCA, Pytheas and ByoPiC products.

\section*{Data availability}
LoTSS DR2 Radio-optical crossmatch catalogue (\citealt{Hardcastle_2023}) is publically available at \url{https://lofar-surveys.org/dr2_release.html}. Data product of the DESI Legacy Imaging Surveys (\citealt{Dey_2019}) is available at \url{https://www.legacysurvey.org/}. The catalogue of SDSS cosmic filaments (\citealt{Malavasi_2020}) can be downloaded at \url{https://l3s.osups.universite-paris-saclay.fr/cosfil.html}.




\bibliographystyle{mnras}
\bibliography{references} 

\end{document}